\documentclass[preprint,showpacs,floatfix]{revtex4}
\usepackage{amssymb}
\usepackage{amsmath}
\usepackage[dvips]{graphicx}

\def\be{\begin{eqnarray} &&}
\def\nonu{\nonumber \\ &&}
\def\ee{\end{eqnarray}}

\def\bew{\begin{widetext}}
\def\ew{\end{widetext}}

\newcommand{\mbf}[1]{\mathbf{#1}}

\def\Dslash{\raise.15ex\hbox{/}\kern-.7em D}
\def\Pslash{\raise.15ex\hbox{/}\kern-.7em P}

\usepackage{color}

\renewcommand{\bar}[1]{\overline{#1}}

\begin{document}
\title{Quantitative studies of the homogeneous Bethe-Salpeter Equation 
in  Minkowski space}
\author{}
\author{Tobias   Frederico$^a$, Giovanni Salm\`e$^b$ and   Michele Viviani$^c$
 }
\affiliation{$^a$ Dep. de F\'\i sica, Instituto Tecnol\'ogico de Aeron\'autica,
DCTA, 12.228-900 S\~ao Jos\'e dos
Campos, S\~ao Paulo, Brazil\\
$^b$Istituto  Nazionale di Fisica Nucleare, Sezione di Roma, P.le A. Moro 2,
 I-00185 Roma, Italy \\
$^c$Istituto  Nazionale di Fisica Nucleare, Sezione di Pisa,
Largo Pontecorvo 3, 56100, Pisa, Italy }
\begin{abstract}
The Bethe-Salpeter Equation for a bound system, composed by two massive scalars
exchanging a massive scalar, is quantitatively investigated in ladder
approximation, within the Nakanishi 
integral representation 
approach. For the S-wave case, numerical solutions with a  form  inspired by the 
Nakanishi integral representation, have been calculated. The needed Nakanishi
weight functions have been evaluated by solving   two different eigenequations, 
 obtained directly   from the Bethe-Salpeter
equation  applying the  Light-Front projection technique.    
A remarkable agreement, 
in particular for
the eigenvalues, has been achieved,  numerically confirming  that the 
Nakanishi uniqueness theorem for the weight functions, demonstrated in the 
context 
of the perturbative
analysis of the multi-leg transition amplitudes
and playing a basic role 
in suggesting one of the two adopted eigenequations, can be extended to a non
perturbative realm.
 The detailed,  quantitative studies are completed by presenting both   
 probabilities and Light-Front
 momentum distributions  for the valence component of the bound state.

\end{abstract}%

 \maketitle



\section{Introduction}\label{intr}

 Solving  the Bethe-Salpeter Equation (BSE)\cite{SB_PR84_51} 
 in   Minkowski space, even 
 for scalar theories, is still a challenging 
 problem, and  not  too
 many numerical
 investigations, 
  able to address the issue, can be found in the literature.  Seemingly, 
  one  of the most  
 effective tool,
 for facing with such a task,  with high numerical accuracy 
 (see, e.g., Refs. \cite{KW1,KW,dse,sauli,carbonell1,carbonell2,carbonell3} for
 an illustration of actual  calculations), 
 is represented by 
 the  so-called perturbation theory
integral representation (PTIR) of the multi-leg transition amplitudes, proposed 
by N. Nakanishi in the sixties (see, e.g. 
\cite{nak63,nak71}). Such an approach originates from the parametric formula 
for
 Feynman integrals and 
leads to a  spectral 
representation of any multi-leg transition amplitude, expressed through 
 an
infinite series of Feynman diagrams. Then, a transition amplitude 
can be written as   a suitable 
 folding
 of a  non singular weight function,  the so-called Nakanishi weight function, 
 divided 
 by a denominator 
 containing the analytic structure
 of the amplitude. In particular, the PTIR  of the three-leg amplitude, i.e. the PTIR vertex function
   (related to the BS amplitude through the inverse of the constituent
   propagators), plays a basic role 
 in  the quest of  physically-motivated, actual  solutions of the BSE 
 in Minkowski space. This is shown 
 by  the   nice results  in Refs 
   \cite{KW1,KW,dse,sauli,carbonell1,carbonell2,carbonell3}, where a wide range
   of i) systems (bosonic and fermionic), ii) approximated kernels (ladder and  
   cross-ladder) and iii) constituent propagators (free and dressed) has been 
   explored.  Loosely speaking, applying PTIR in order to 
   solve the BSE can be seen as a generalization 
of the approach proposed by Wick and Cutkosky
\cite{WICK_54,CUTK_54} for obtaining explicit solutions of  the 
scalar-scalar BSE, but  with a massless-scalar exchange. 
 
 Among the attractive features of PTIR, one has i)  the 
 dependence of the non singular 
 weight functions    upon  real 
 variables, whose  number is related to the number of 
  independent invariants of the problem, and ii)
  the explicit analytic structure of the
 transition amplitude, that allows one  to perform
 analytic
 integrations, when  requested (indeed,  this will be the case). These properties
 have been exploited in order to obtain equations for the Nakanishi weight
 function starting  from the BSE. In particular,
  (cf Refs.
    \cite{KW1,KW,dse,sauli,carbonell1,carbonell2,carbonell3}) 
   one can single out two different equations for determining
   the Nakanishi weight function, but
    both of them share the first step: one assumes a form like 
  the PTIR vertex for 
  the BS amplitude,  and puts such an expression in the BSE. Then, one can
  proceed by invoking the Nakanishi {\em uniqueness theorem} for the
  weight functions \cite{nak71} and obtains the  eigenequation for 
 the Nakanishi weight function adopted in Refs.
 \cite{KW1,KW,dse,sauli}, where the  ladder
 approximation of the BS kernel has been assumed and  some 
 elaborations, like either free or 
 dressed 
 constituent propagators have been proposed.
   Indeed, the uniqueness theorem was  
  demonstrated within the perturbative
   analysis of the transition
  amplitudes, and therefore one could ask if and to what extent this can be
  applied in a  non
  perturbative framework: this is a first question we have addressed in the
  present paper. Moreover, in Refs. \cite{KW1,KW,dse,sauli}, the truncated kernel 
   was
  elaborated by performing  the needed analytical integrations  using the
  standard Minkowski variables.
  Differently, in Ref. \cite{carbonell1,carbonell2,carbonell3}, an 
  explicitly-covariant Light-front (LF) 
  approach \cite{cdkm}, with  the set of LF variables, $k^\pm=k^0\pm k^3$ and 
  ${\bf
k}_\perp\equiv \{k^1,k^2\}$, was
 adopted for determining another integral equation for the Nakanishi weight 
  function.
  In particular, through an exact relation based on a suitable analytic 
  integration of the Nakanishi  representation of the BS amplitude, one can
  extract the so-called {\em valence component} of the interacting-system state (see
  also Ref. \cite{Yabuki,Burov}), i.e.
  the first contribution to the Fock expansion of the state. Applying the same
  integration on both sides of the BSE
  one gets an integral equation for the weight function, that has  on the lhs 
  the
  valence component and 
  on the rhs the Nakanishi function  combined with a proper kernel (obtained 
  from the BS 4D kernel). It should be pointed out
  that such an  integral equation is a generalized eigenequation and therefore
  substantially different from the one of Refs. \cite{KW1,KW,dse,sauli}, that
  is a true eigenequation, once the ladder approximation is adopted.

  A formal investigation for establishing a direct bridge 
 between  the above described
  approaches,  also extending the analysis from  the homogeneous 
  (bound states) to
  inhomogeneous (scattering states) ladder BSE, was performed in Ref.
  \cite{FSV}. To accomplish this,  a non-explicitly-covariant LF framework
  was adopted  together with a LF projection technique (see,  Refs.
   \cite{sales00,sales01,hierareq,adnei07,adnei08} for details). In particular the integration over the LF variable $k^-$ was
  exploited  for obtaining the valence component from the BS amplitude, 
  arriving
  at
  the same generalized eigenequation of Ref. \cite{carbonell1} for the
  Nakanishi weight function,  but deduced within  the explicitly covariant LF
  framework (see Ref. \cite{cdkm}).
  Moreover, by a suitable elaboration of the kernel present  in the
  generalized eigenequation one become ready for applying the Nakanishi
  uniqueness theorem \cite{nak71}. It should be pointed out that 
  all the formal developments in Ref. \cite{FSV}  benefit 
  from the well-known  virtue of the 
  LF variables  to make simpler the analytical integrations  
  (see Ref. \cite{Sawicki} 
  for an elementary introduction to the issue).
  
 Aim of the present work is the numerical investigation of the above mentioned
 eigenequations (i.e. with and  without the application of the
 uniqueness theorem), in order to evaluate the Nakanishi weight functions 
 of  BS
 amplitudes, solutions of a ladder   BSE  for a S-wave bound system, composed by two
 massive scalars,
 interacting through the exchange of a massive scalar.
 In particular, for both eigenequations we have calculated 
 i) eigenvalues and eigenfunctions,  corresponding to 
 binding energies and  masses of the exchanged scalar of Refs. 
 \cite{KW} and
 \cite{carbonell1}, and  ii) valence
 probabilities and  LF distributions 
 in  both the longitudinal-momentum fraction, $\xi$,
  and the transverse momentum, $|{\bf k}_\perp|$,
   that notably can be evaluated once the Nakanishi
 weight functions are determined.
The comparison between the numerical results  obtained from the two 
eigenequations allows
us to check to which extent  the uniqueness theorem of
the Nakanishi weight function is valid, and to assess the reliability of quantities,
like valence probabilities and LF distribution, quite relevant in the
phenomenological studies of interacting, relativistic  systems.
 
 The  paper is organized as follows. In Sect. \ref{sbslf}, 
the general formalism of the BSE onto the null plane is introduced, as well as the valence component of the BS
amplitude. 
In Sect. \ref{slad}, the kernel of the ladder BSE is recast in a form suitable for  applying the uniqueness
theorem  by Nakanishi. In Sect. \ref{slf}, the  LF momentum distributions are defined.
In Sect. \ref{ris}, the numerical results are discussed. Finally, in Sect. \ref{concl}, the conclusions are
drawn.

\section{The homogeneous Bethe-Salpeter equation onto the null plane}
\label{sbslf}
 In this Section, the general formalism adopted for obtaining the
 eigenequations for the Nakanishi weight-functions, within the LF framework of
 Ref. \cite{FSV}, is quickly reviewed, in
 order to have the full matter under control, and proceed in the following 
 Sections 
 to
 the numerical analysis. Moreover, it is illustrated  (Appendix A contains the
 details)
  a shorter way to deduce  the
 eigenequation based on the  uniqueness theorem from the one based on the LF
 valence wave function. It should be pointed out that the BSE we have
 considered does not
 contain neither self-energy  nor vertex corrections, but  it worth mentioning
  that
 one could rely upon  a Dyson-Schwinger framework for dressing the constituent
 propagators
 (see,e.g., Ref. \cite{dse}).
 
 Let us start recalling that the BS amplitude  for a bound state fulfills the 
 following BSE
   \be
\Phi_b(k,p)= G_0^{(12)}(k,p)
~
\int \frac{d^4k^\prime}{(2\pi)^4}i~{\cal K}(k,k^\prime,p)\Phi_b(k^\prime,p),
\label{BSE}\ee
where  $p=p_1+p_2$ is the total momentum of the interacting system 
with total mass $p^2=M^2$, $k=(p_1-p_2)/2$  the relative momentum and 
$i~{\cal K}$  the interaction kernel, that
contains all the irreducible 
 diagrams.  As mentioned above the self-energy is disregarded, and therefore 
one has to consider the free propagator of the two constituents
$G_0^{(12)}$,   given by
\be\label{G0}
G_0^{(12)}(k,p) =G_0^{(1)}G_0^{(2)}
=\frac{i}{(\frac{p}{2}+ k)^2-m^2+i\epsilon}~~~
\frac{i}{(\frac{p}{2}-k)^2-m^2+i\epsilon} \ .
\ee
In what follows, we look for S-wave solutions of Eq. (\ref{BSE}), that can be written
as the PTIR vertex function, i.e. a proper folding of a non singular weight function, that
depends upon real variables, and a factor that contains the analytic structure
 (see also Refs. \cite{KW1,KW,dse,carbonell1,carbonell2,carbonell3,FSV}), namely 
 \be\label{bsint}
\Phi_b(k,p)=
~i ~\int_{-1}^1dz'\int_0^{\infty}d\gamma'
~
{g_b(\gamma',z';\kappa^2)\over\left[\gamma'+\kappa^2-{k}^2-p\cdot k
z'-i\epsilon\right]^{2+n}}
\label{naka1}\ee
where  $g_b(\gamma',z';\kappa^2)$ is the Nakanishi weight function, 
and 
$\kappa^2$ is defined
  by 
 \be\label{kappa2}
\kappa^2 = m^2- {M^2\over 4}.
\ee
with $m$ the constituent mass. Notice that, by definition, one has 
$\kappa^2 > 0$  for bound states, while $\kappa^2 <0$ for scattering states. 
The power $n$ in  the denominator of Eq. (\ref{naka1}) 
can be any value $n\geq 1$. The minimal value $n=1$ ensures the convergence of
the 4D integral, and in what follows we adopt this choice, as in    
 Refs. \cite{KW,carbonell1}.
Increasing the value of  $n$ should produce a   solution for
$g_b(\gamma',z';\kappa^2)$ more and more soft
\cite{KW,carbonell5}. The factor of $2$  in the exponent of the denominator of
Eq. (\ref{naka1}) comes from the fact we are dealing with the
BS amplitude and not with the vertex function.

It is worth noting that  the dependence upon $z'$ of $g_b(\gamma',z';\kappa^2)$ is even
 as expected by the symmetry property of the BS amplitude for the
two-scalar system. As a matter of fact, when the exchange between  the two 
particles is performed,  the scalar product $ k \cdot p$ in the denominator in Eq. 
(\ref{naka1}) changes sign. In order to recover the expected symmetry of the 
BS amplitude, the Nakanishi
weight function must be even in $z'$. Moreover, as pointed out in Refs. 
\cite{nak63,nak69,nak88,KW},
  $z$-odd 
$g_b(\gamma',z';\kappa^2)$ functions  correspond to odd-parity BS amplitudes
 with respect to
the change $k^0\to -k^0$ (recall that in the rest frame $p\cdot k =M k^0$). 
It turns out (see Ref. \cite{nak63,nak69,nak88} for
more details) that such BS amplitudes have 
 negative norm.

 As  is well known (see, e.g., Refs.\cite{Brodrev,FSV} for the non-explicitly-covariant 
 LF  approach and Refs. \cite{cdkm,carbonell1} for the explicitly-covariant 
 case), 
 one can obtain the valence component of the interacting state from the
 corresponding BS amplitude, through the suitable analytic integration, namely
 the integration over $k^-$.
  Once the expansion  of the interacting state on a
 Fock basis is introduced,   the valence component corresponds to the 
 contribution with the
 minimal number of constituents, that in the present case amounts to  two scalars.
 For the sake of clarity, it is useful to briefly recall the above mentioned 
 procedure, within the LF framework adopted in our previous work
 \cite{FSV}, since in the
 following Sections the valence probability and the LF distributions will be
 discussed and numerically evaluated.
 
 In the Fock space one can introduce  the   completeness given by
 \be
 \sum_{n \ge 2}~ \int \left [d^3 \tilde q_i \right]
 ~\big\vert n; \tilde q_i\bigr\rangle~
\bigl\langle \tilde q_i ; n \big\vert= ~{\cal I}
\ee
with ${\cal I}$ the identity in  the Fock space,  $\tilde q_i\equiv \{q^+_i, \mbf{q}_{i\perp}\}$
the LF three-momenta, and  
\be \big\vert n; \tilde q_i\bigr\rangle= (2 \pi)^{3n/2} ~{1 \over \sqrt{n!}}
\sqrt{ 2q^+ _1}
 \dots ~ \dots \sqrt{2q^+ _n}~a^\dagger_{\tilde q_1} \dots~ \dots a^\dagger_{\tilde q_n}~\big\vert  0\bigr\rangle 
\label{focks} \ee  The normalization for the single-particle free state is  
$\bigl\langle \tilde q' \big\vert\tilde q \bigr\rangle = 
2 q^+ (2 \pi)^3\delta^3
\bigl(\tilde q ' - \tilde q\bigr)$, that leads to the standard LF   
phase space, viz
\be
\int \left [d^3 \tilde q_i \right] =\prod_{i=1}^n \int {d^3 \tilde q_i \over  2 q_i^+ (2 \pi)^3}
\ee
The free Fock states in Eq. (\ref{focks}) have the following orthonormalization
\be
\langle \tilde q'_i; n'\big\vert n; \tilde q_i\bigr\rangle=~{1 \over n!}~\delta_{n,n'}~\sum_{[j_1\dots j_n]_{perm}}~~\prod_{i=1}^n ~2 ~q^+_i~(2\pi)^3~ 
\delta^3
\bigl(\tilde q'_i - \tilde q_{j_i}\bigr)
\ee
where the sum has to be performed over all the $n!$ permutations of $1 \dots \dots n$, as shortly indicated by $[j_1\dots j_n]_{perm}$.
Then,  the interacting state can be expanded as follows
(see, e.g.,  Ref. \cite{Brodrev})
\be
\left\vert \tilde p; \Psi_{int} \right\rangle = 2~(2\pi)^3~\sum_{n\ge 2}\int \big[d \xi_i\big] \left[d^2
\mbf{k}_{i\perp }\right]\,
\psi_{n/p}(\xi_i,\mbf{k}_{i\perp })~ \bigl\vert n; \xi_i
p^+,~  \mbf{k}_{i\perp } + \xi_i \mbf{p}_{\! \perp}\! \bigr\rangle,
\label{LFstate}
\ee
where i) $\bigl\vert n;~ \xi_i
p^+,~  \mbf{k}_{i\perp } + \xi_i \mbf{p}_{\! \perp} \! \bigr\rangle$ is the Fock
state
 with $n$ particles; and  the variables $\tilde q_i$ have been expressed  in terms of 
 the intrinsic variables, 
 $\{\xi_i, \mbf{k}_{i\perp }\}$ 
as follows: $q^+_i=\xi_i p^+$ and
$\mbf{q}_{i \perp}= \mbf{k}_{i\perp }+ \xi_i ~\mbf{p}_{\perp }$; ii) 
$\psi_{n/p}(\xi_i,\mbf{k}_{i\perp })$ are the so-called LF wave
 functions, that allow one to describe the intrinsic dynamics  and are related
 to the  overlap $\bigl\langle
n;~\xi_i p^+,~\mbf{k}_{i\perp } +\xi_i \mbf{p}_{\perp }\big\vert \tilde p; 
\Psi_{int}\bigr\rangle$ 
as discussed below.  
Notice that  
 the global motion and the intrinsic structure have been kept separate in
$\left\vert \tilde p; \Psi_{int} \right\rangle$, given the kinematical nature of
the LF boosts. Finally, let us remind that 
the interacting system is on-mass-shell, i.e. $p^-=(M^2 +
|\mbf{p}_{\! \perp}|^2)/ p^+$, and 
the set of intrinsic variable $\{\xi_i,\mbf{k}_{i\perp }\}$
  satisfy the
 following relations
 \be \sum_{i=1}^n \xi_i=1 \quad \quad \sum_{i=1}^n\mbf{k}_{i\perp} =0\ee
The phase-space factors in Eq. 
(\ref{LFstate}) is given by 
\be
\int \big[d \xi_i\big] \equiv \prod_{i=1}^n \int {d\xi_i\over 2 ~(2\pi) \xi_i } \,\delta
\Bigl(1 - \sum_{j=1}^n \xi_j\Bigr)=~p^+~ \prod_{i=1}^n \int {dq^+_i\over 2q^+_i 
(2\pi)  }~\delta\Bigl(p^+ - \sum_{j=1}^n  q^+_j\Bigr) ,
\nonu
\int \left[d^2 \mbf{k}_{i\perp }\right] \equiv \prod_{i=1}^n \int
{d^2 \mbf{k}_{i\perp }\over (2\pi)^2}
\delta^{2} \Bigl(\sum_{j=1}^n\mbf{k}_{j\perp}\Bigr).
\label{phase}\ee
 Since the intrinsic motion is kinematically separated from the global one, within the LF framework, the overlap $\bigl\langle
n;~\xi_i p^+,~\mbf{k}_{i\perp } +\xi_i \mbf{p}_{\perp }\big\vert \tilde p;; 
\Psi_{int} \bigr\rangle$ can be  trivially factorized into the product of
a momentum-conserving delta function and the  intrinsic LF 
  wave
 function  as follows
\be \label{eq:LFWF}
 \bigl\langle
n;~\xi_i p^+,~\mbf{k}_{i\perp }+ \xi_i \mbf{p}_{\perp }\big\vert \tilde p;\Psi_{int} \bigr\rangle
= 2p^+ (2\pi)^3 \delta^3\Bigl(\tilde p-\sum_{i=1}^n \tilde q_i\Bigr)~\psi_{n/p}(\xi_i, \mbf{k}_{i\perp })=
\nonu=
2 (2\pi)^3 \delta \bigl(1 - \sum_{i=1}^n \xi_i \bigr) \,
\delta^{(2)}
\bigl(\sum_{i=1}^n \mbf{k}_{i\perp} \bigr)~\psi_{n/p}(\xi_i, \mbf{k}_{i\perp })
\ee
  From Eq. (\ref{LFstate}) and  reminding that the CM
plane waves have the standard normalization that can be factorized out,
one can obtain the normalization of the intrinsic state and in turn 
  i) the overall normalization of the 
LF wave functions and ii) the probability of each Fock component. As a matter of fact, one can write
\be
\bigl\langle \tilde p'; \Psi_{int}\big\vert \tilde p;\Psi_{int}\bigr\rangle = 2 p^+~(2 \pi)^3\delta^3
\bigl(\tilde p' - \tilde p\bigr)~\bigl\langle  \Psi_{int}\big\vert \Psi_{int}\bigr\rangle= 
\nonu =
[2 p^+ (2\pi)^3]^2 \sum_{n \ge 2} 
 \int \left[ d^3 \tilde q_i \right] 
 \delta^3\Bigl(\sum_{i=1}^n \tilde q_i- \tilde p\Bigr)~\psi_{n/p}(\xi_i, \mbf{k}_{i\perp })
 \delta^3\Bigl(\sum_{i=1}^n \tilde q_i- \tilde p'\Bigr)~\psi_{n/p'}(\xi_i, \mbf{k}_{i\perp })=
 \nonu= 2p^+ (2\pi)^3 \delta^3\bigl(\tilde p' - \tilde p\bigr)~ 2~(2\pi)^3~\sum_{n\ge 2}\int \big[d \xi_i\big] \left[d^2
\mbf{k}_{i\perp }\right]\,
\big\vert  \psi_{n/p}(\xi_i, \mbf{k}_{i\perp }) \big\vert^2 
  \ee
Then, the  LF wave functions, $\psi_{n/p}(\xi_i, \mbf{k}_{i\perp })$, are normalized according to
\be
\bigl\langle  \Psi_{int}\big\vert \Psi_{int}\bigr\rangle= 2~(2\pi)^3~\sum_{n \ge 2}  \int \big[d \xi_i\big] \left[d^2 \mbf{k}_{i\perp }\right]
\,\left\vert \psi_{n/p}(\xi_i, \mbf{k}_{i\perp }) \right\vert^2 = 1.
\label{LFWFnorm}
\ee
This equation clearly shows the physical content associated to the LF wave
functions: $\left\vert \psi_{n/p}(\xi_i, \mbf{k}_{i\perp }) \right\vert^2$
yields the probability distributions to find $n$ constituents with
intrinsic coordinates $\{ \xi_i, \mbf{k}_{i\perp }\}$ inside the 
interacting-system state. In view of this, it should be pointed out  the
basic role played by  LF wave functions in extracting the probability content hidden inside the BS amplitude.
Notice that a factor $2~(2\pi)^3$ is missing in the corresponding equation 
(i.e. Eq.(15)) of
 Ref. \cite{FSV}.

In particular, the probability to find the valence 
component in the bound state (see Sect. \ref{slf})
is given by 
\be
N_2= 2~(2\pi)^3~ \int {d\xi_1\over 2 ~(2\pi) \xi_1 } \int {d\xi_2\over 2 ~(2\pi) \xi_2 } ~\delta(1 -\xi_1-\xi_2)~\times
\nonu
\int
{d^2 \mbf{k}_{1\perp }\over (2\pi)^2}\int
{d^2 \mbf{k}_{2\perp }\over (2\pi)^2}~ \delta^{2} \Bigl(\mbf{k}_{1\perp}+\mbf{k}_{2\perp}\Bigr)~
\left\vert \psi_{n=2/p}(\xi_1, \mbf{k}_{1\perp } ) \right\vert^2=\nonu
=  {1 \over (2 \pi)^3}~\int {d\xi \over 2 ~ \xi (1-\xi)} 
\int
d^2 \mbf{k}_{\perp }~
\left\vert \psi_{n=2/p}(\xi, \mbf{k}_{\perp } ) \right\vert^2
\label{N2}\ee
where the notation has been simplified, putting $\xi=\xi_1$ and 
$\mbf{k}_{\perp }=\mbf{k}_{1\perp }$.
 In general, the probability $N_n$
of the $n$-th Fock component can be evaluated through the corresponding LF wave
function.
  
 In general, the valence wave function can be obtained by integrating 
 the BS
 amplitude $\Phi_b(k,p)$ over $k^-$ (see \cite{FSV} for details). Once we assume the
 expression for
 the  BS amplitude  suggested by the PTIR approach \cite{nak71},
  we get  (see also \cite{carbonell1})
\be \label{val1}
\psi_{n=2/p}(\xi,k_\perp)=~{p^+\over \sqrt{2}}~\xi~(1-\xi)
\int {dk^- \over 2 \pi}
\Phi_b(k,p)=\nonu
=~{1\over \sqrt{2}}\xi~(1-\xi)~\int_0^{\infty}d\gamma'~\frac{
g_b(\gamma',1-2\xi;\kappa^2)}
{[\gamma'+k_\perp^2 +\kappa^2+\left(2\xi-1\right)^2 {M^2\over 4}-i\epsilon]^2}.
\ee
where the integration over $k^-$ leads to fix the value
of the variable $z'$ in Eq. (\ref{naka1}) to $1-2\xi$. The factor $ 1/\sqrt{2}$
comes from the normalization of the Fock state with $n=2$, given the statistics
property. 

From Eq. (\ref{val1}) and the physically-motivated request that the
density in the transverse variable ${\bf b}_\perp$, conjugated to 
${\bf k}_\perp$, be finite for $|{\bf b}_\perp|=0$, one can deduce that 
$g_b(\gamma',1-2\xi;\kappa^2)$ must vanish for $\gamma' \to \infty$. As a
matter of fact, one has  
\be
\tilde\psi_{n=2/p}(\xi,b_\perp)= \int {d {\bf k}_\perp \over (2 \pi)^2}~ 
e^{i{\bf k}_\perp \cdot {\bf b}_\perp}~ \psi_{n=2/p}(\xi,k_\perp)
\label{valbt}\ee
and
 \be
\tilde\psi_{n=2/p}(\xi,b_\perp=0)= \pi {1\over \sqrt{2}}\xi~(1-\xi)\int_0^\infty {d  k^2_\perp \over (2 \pi)^2}~ 
 \int_0^{\infty}d\gamma'~\frac{
g_b(\gamma',1-2\xi;\kappa^2)}
{[\gamma'+k_\perp^2 +\kappa^2+\left(2\xi-1\right)^2 {M^2\over 4}-i\epsilon]^2}=\nonu=
  {1\over 4\pi \sqrt{2}}\xi~(1-\xi)\int_0^{\infty}d\gamma'~\frac{
g_b(\gamma',1-2\xi;\kappa^2)}
{[\gamma'+\kappa^2+\left(2\xi-1\right)^2 {M^2\over 4}-i\epsilon]}
\label{valbt1}\ee
If the transverse density at the origin, i.e. 
$|\tilde\psi_{n=2/p}(\xi,b_\perp=0)|^2$,  is finite, one
can immediately  realize  the needed fall-off of $g_b(\gamma',1-2\xi;\kappa^2)$. 
Notice that   the denominator is always positive for a bound state.
By introducing the  variables ($\gamma,z$), as in  Ref.
\cite{carbonell1},
\be \gamma=k_{\perp}^2 \quad \quad   1\geq z=1-2\xi\geq -1 ~~~~.\ee
 one can rewrite the valence wave function as follows
\be \label{val2}
\psi_{n=2/p}(z,\gamma)
={(1-z^2)\over 4 \sqrt{2}}~\int_0^{\infty}d\gamma'~\frac{
g_b(\gamma',z;\kappa^2)}
{[\gamma'+\gamma +z^2m^2+(1-z^2)\kappa^2-i\epsilon]^2}
\ee
The announced integral equation for the Nakanishi weight function,
$g_b(\gamma, z;\kappa^2)$, is obtained by inserting (\ref{bsint}) in both
sides of the LF-projected BS equation (see also
\cite{sales00,sales01,hierareq,adnei07,adnei08}), viz 
\be\label{BST} \int {dk^-\over 2 \pi}~\Phi_b(k,p)= \int {dk^- \over
2 \pi}~G_0^{(12)}(k,p) \int\frac{d^4k^\prime}{(2\pi)^4} i~{\cal
K}(k,k^\prime,p)\Phi_b(k^\prime,p), \ee 
Then,  one gets \cite{FSV} (see Ref. \cite{carbonell1} for the corresponding
elaboration within the explicitly-covariant LF
framework)
\be
\int_0^{\infty}d\gamma'~\frac{g_b(\gamma',z;\kappa^2)}{[\gamma'+\gamma
+z^2 m^2+(1-z^2)\kappa^2-i\epsilon]^2} =\nonu=
\int_0^{\infty}d\gamma'\int_{-1}^{1}dz'\;V^{LF}_b(\gamma,z;\gamma',z')
g_b(\gamma',z';\kappa^2).
\label{ptireq}\ee
where the new kernel $V^{LF}_b$, that we call Nakanishi kernel for the sake of
brevity, is
related to the BS kernel, $i{\cal K}$, in Eq. (\ref{BSE}), as follows
\be\label{V}
V^{LF}_b(\gamma,z;\gamma',z')=~i
p^+~\int_{-\infty}^{\infty}{d k^- \over 2\pi}~G_0^{(12)}(k,p)
\int \frac{d^4k'}{(2\pi)^4}\frac{i{\cal K}(k,k',p)}
{\left[{k'}^2+p\cdot k' z'-\gamma'-\kappa^2+i\epsilon\right]^3}
\label{vbou}\ee

A different equation for $g_b(\gamma, z;\kappa^2)$ can be obtained, still
starting from the LF-projected BSE (\ref{BST}), if one takes into account i) the
uniqueness of the Nakanishi weight function, as ensured 
by a theorem  in Ref. \cite{nak71} and ii) the PTIR expressions for both the BS
amplitude and  the BS kernel, i.e. a four-leg transition amplitude (see, e.g., 
\cite{KW,FSV}, for the actual PTIR of the off-shell T-matrix). Then, 
in place of  Eq. (\ref{ptireq}),  one
could  write   the following eigenequation (see 
\cite{nak64,nak642,KW1,KW,dse,sauli,FSV} for the ladder case)
\be
g_b(\gamma,z;\kappa^2)=~\int_{0}^{\infty}d\gamma'\int_{-1}^{1}dz'\;
{\cal V}_b(\gamma,z;\gamma',z';\kappa^2)
g_b(\gamma',z';\kappa^2)
\label{uniq}\ee
Within the PTIR framework, it is very important to notice  that both Eqs. (\ref{ptireq}) and (\ref{uniq}) are 
equivalent to the initial BSE (\ref{BSE}), if    the
uniqueness theorem holds.
Once the weight function
$g_b(\gamma',z;\kappa^2)$ is known, then one can fully reconstruct, in
Minkowski space (see Eq. (\ref{bsint})),  the   BS amplitude, {\em that belongs to the
class of physically acceptable solutions} (with positive norm and suitable for
an investigation within a S-matrix framework).
Moreover, it is not surprising that through the information stored in 
the   valence
component one can map the full BS amplitude, since the whole,   rich content 
of the BS 
amplitude can be transferred to the LF kernel, i.e. the kernel projected 
onto the null
plane. This result is quite general and
holds both in  perturbative and non perturbative regimes, 
 and, even more, for both bound and
 scattering states
 \cite{sales00,sales01,hierareq,adnei07,adnei08}. 
 
\section{LF Nakanishi Kernel in Ladder Approximation} 
\label{slad} 
At the present stage, our numerical investigation is restricted to the ladder
approximation of the BSE, where the BS kernel  is given by
\be
i{\cal K}^{(Ld)} (k,k')={i (-ig)^2 \over (k-k')^2 -\mu^2 +i\epsilon}
\ee
with $\mu$  the mass of the exchanged scalar. 
 Explicit expressions for both   ladder and  cross-ladder
approximations of $V^{LF}_b(\gamma,z;\gamma',z')$, obtained within the covariant
LF framework, 
can be found in Refs. \cite{carbonell1,carbonell2}.

In Ref. \cite{FSV},  where  a non-explicitly-covariant LF framework was chosen,
 the scattering case was analyzed in great detail, and the
ladder approximation of 
  the Nakanishi kernel in  the S-wave bound state (see  Eq. (\ref{ptireq})), $V_b^{(Ld)}$, was  obtained through a
  proper limit of the scattering kernel. In what follows, a more direct and
  simple way to obtain  $V_b^{(Ld)}$, is presented (see  Appendix \ref{nkern}
  for more  details), eventually achieving 
  an   expression  
suitable for exploiting
 the uniqueness theorem of   the Nakanishi weight function \cite{nak71}.
As to the  numerical calculations, the results evaluated with
our LF  approach and the ones shown in Refs. \cite{KW} and \cite{carbonell1}, 
are compared   in Sect. \ref{ris}.

In a reference frame where $\mbf{p}_\perp=0$ and $p^{\pm}=M$, 
the ladder approximation of $V^{LF}_b(\gamma,z;\gamma',z')$, to be 
inserted
in the integral equation (\ref{ptireq}), is written  as follows
 (see Appendix \ref{nkern} for details)
\be
V^{(Ld)}_b(\gamma,z;\gamma',z')=
~-~g^2p^+
\int \frac{d^4k''}{(2\pi)^4}\frac{1}
{\left[{k''}^2+p\cdot k'' z'-\gamma'-\kappa^2+i\epsilon\right]^3}
~\times \nonu \int_{-\infty}^{\infty}{d k^- \over 2\pi}~
{1 \over \left[(\frac{p}{2}+k)^2-m^2+i\epsilon\right]}
~ {1\over \left[(\frac{p}{2}-k)^2-m^2+i\epsilon\right]}{1\over (k-k'')^2
-\mu^2 +i\epsilon}=
\nonu=  ~- ~
~{g^2 \over 2(4\pi)^2}~\int^\infty_{-\infty}
 d\gamma'' ~
{\theta(\gamma'') \over
\left[\gamma+\gamma'' +z^2 m^2+ \kappa^2(1-z^2)  -i\epsilon
\right]^2}~\times \nonu
 \left [{(1+z)\over (1+\zeta')}
~\theta (\zeta'-z)~ h'(\gamma'',z;\gamma',\zeta',\mu^2)
+{(1-z)\over (1-\zeta')}
~\theta
(z-\zeta')~ h'(\gamma'',-z;\gamma',-\zeta',\mu^2)\right]
\label{nkern1}\ee
where
\be
h'(\gamma'',z;\gamma',\zeta',\mu^2) =
\theta \left[\gamma'' { (1+\zeta')\over (1+z)}  - \gamma'-\mu^2- 2\mu
\sqrt{{\zeta'}^{2} \frac{M^2}{4} + \kappa^2+\gamma'}~\right]
 \nonu \times
\left[- {{\cal B}_b(z,\zeta',\gamma',\gamma'',\mu^2) \over
{\cal A}_b(\zeta',\gamma',\kappa^2)~\Delta(z,\zeta',\gamma', \gamma'',\kappa^2,
\mu^2 ) }~{1\over\gamma''} \right. 
\nonu \left. + { (1+\zeta')\over (1+z)}
 \int_{y_-}^{y_+}{dy}~{y^2 \over \left[ {y}^2{\cal A}_b(\zeta',\gamma',\kappa^2 )+ y(\mu^2 +\gamma')+\mu^2
\right ]^2}\right]
\nonu
-{ (1+\zeta')\over (1+z)}
\int_0^{\infty}{dy}~{y^2 \over \left[ {y}^2{\cal A}_b(\zeta',\gamma',\kappa^2 )+ y(\mu^2 +\gamma')+\mu^2
\right ]^2} 
\label{hprime}
\ee
with
\be
{\cal A}_b(\zeta',\gamma',\kappa^2 )={\zeta'}^{2} \frac{M^2}{4} +
\kappa^2+\gamma'~\geq ~0
\nonu
{\cal B}_b(z,\zeta',\gamma', \gamma'',\mu^2 )=
\mu^2+\gamma' -\gamma''
{ (1+\zeta')\over (1+z)}~\leq 0
\nonu
\Delta^2(z,\zeta',\gamma', \gamma'',\kappa^2,\mu^2 )=
{\cal B}_b^2(z,\zeta',\gamma',\gamma'',\mu^2)
- 4 \mu^2~ {\cal A}_b(\zeta',\gamma',\kappa^2)~\geq ~0
\nonu
y_\pm=
{1 \over 2{\cal A}_b(\zeta',\gamma',\kappa^2)} 
 \left[ -{\cal B}_b(z,\zeta',\gamma',\gamma'',\mu^2)
 \pm \Delta(z,\zeta',\gamma', \gamma'',\kappa^2,\mu^2 )\right]
 \label{nkern2}
 \ee
 It is relevant for what follows that for  $z\to (-1)$: i) the theta function does not anymore yields a constraint; ii)  the function ${\cal B}_b \to -\infty$ and
 iii) the two integrals on $y$ cancels each other.  Then, one gets   $$
{(1+z)\over (1+\zeta')} h'(\gamma'',z;\gamma',\zeta',\mu^2) \to
{(1+z)\over (1+\zeta')} {1\over   (\gamma '' ~{\cal A}_b
(\zeta',\gamma',\kappa^2 ))}
\to 0$$
An analogous result can be obtained for $z\to 1$ when the term containing 
$h'(\gamma'',-z;\gamma',\zeta',\mu^2) $ is considered.

Notably, in Eq. (\ref{nkern1}) the denominator $ 1/[\gamma+\gamma'' +z^2 m^2+ \kappa^2(1-z^2)  -i\epsilon
]^2$ has been factored out, making possible the application of the 
uniqueness theorem to the ladder approximation of Eq. (\ref{ptireq}), that reads
\be
\int_0^{\infty}d\gamma'~\frac{g^{(Ld)}_b(\gamma',z;\kappa^2)}{[\gamma'+\gamma
+z^2 m^2+(1-z^2)\kappa^2-i\epsilon]^2} =\nonu =
\int_0^{\infty}d\gamma'\int_{-1}^{1}dz'\;V^{(Ld)}_b(\gamma,z;\gamma',z')
g^{(Ld)}_b(\gamma',z';\kappa^2).
\label{ptirl}\ee
As a matter of fact, one can rewrite Eq. (\ref{ptirl})
as follows
\be
\int_0^{\infty}d\gamma'~\frac{g^{(Ld)}_b(\gamma',z;\kappa^2)}{[\gamma'+\gamma
+z^2 m^2+(1-z^2)\kappa^2-i\epsilon]^2} =\nonu=
{g^2 \over 2(4\pi)^2}~\int_0^{\infty}d\gamma' \int_{-1}^{1}d\zeta'\;
g^{(Ld)}_b(\gamma',\zeta';\kappa^2)
\nonu 
\int^\infty_{0}
 d\gamma'' ~
{1 \over
\left[\gamma+\gamma'' +z^2 m^2+ \kappa^2(1-z^2)  -i\epsilon
\right]^2}~\times \nonu
 \left [{(1+z)\over (1+\zeta')}
~\theta (\zeta'-z)~ h'(\gamma'',z;\gamma',\zeta',\mu^2)
+{(1-z)\over (1-\zeta')}
~\theta
(z-\zeta')~ h'(\gamma'',-z;\gamma',-\zeta',\mu^2)\right]
\label{newbound}\ee
and,  after changing the name of the integration variable
in the lhs, one gets
\be
\int_0^{\infty}d\gamma''~\frac{1}{[\gamma''+\gamma
+z^2 m^2+(1-z^2)\kappa^2-i\epsilon]^2}~\times
\nonu \left [g^{(Ld)}_b(\gamma'',z;\kappa^2)- 
{g^2 \over 2(4\pi)^2}~\int_0^{\infty}d\gamma' \int_{-1}^{1}d\zeta'\;
g^{(Ld)}_b(\gamma',\zeta';\kappa^2)
~\right. \times \nonu \left.
 \left ({(1+z)\over (1+\zeta')}
~\theta (\zeta'-z)~ h'(\gamma'',z;\gamma',\zeta',\mu^2)
+{(1-z)\over (1-\zeta')}
~\theta
(z-\zeta')~ h'(\gamma'',-z;\gamma',-\zeta',\mu^2)\right)\right]=0
\nonu
\label{newbound1}\ee
 If the uniqueness theorem holds, then  the ladder approximation 
 of Eq. (\ref{uniq}) reads
\be
g^{(Ld)}_b(\gamma'',z;\kappa^2)=
\int_{0}^{\infty}d\gamma'\int_{-1}^{1}d\zeta'\;
{\cal V}^{(Ld)}_b(\gamma'',z;\gamma',\zeta';\kappa^2)
=\nonu=~{\alpha ~m^2\over 2\pi}~
~\int_0^{\infty}d\gamma' \int_{-1}^{1}d\zeta'\;
g^{(Ld)}_b(\gamma',\zeta';\kappa^2)
~\times \nonu 
 \left [{(1+z)\over (1+\zeta')}
~\theta (\zeta'-z)~ h'(\gamma'',z;\gamma',\zeta',\mu^2)
+{(1-z)\over (1-\zeta')}
~\theta
(z-\zeta')~ h'(\gamma'',-z;\gamma',-\zeta',\mu^2)\right]
\label{newbound2}\ee
where $$\alpha={1\over 16\pi}~\left({g \over m}\right)^2 ~~~~~.$$
is an adimensional quantity, since in our model Lagrangian,
 ${\cal L}_{int}=g~ \Phi^\dagger_a \Phi_a
\phi_b$, the coupling constant $g$  has the dimension of a mass (as it must be for a $\phi^3$
theory).
It is worth noting that the kernel between  the square brackets is symmetric with respect to the
transformation $\{z,\zeta'\} \to \{-z,-\zeta'\}$. Moreover 
$g_b(\gamma,z=\pm 1;\kappa^2)=0$,  given the presence of the theta functions and
the vanishing value of 
$${(1\pm z)\over (1\pm \zeta')} h'(\gamma'',\pm z;\gamma', \pm \zeta',\mu^2)$$
for $z \to~\mp 1$ as discussed below Eq. (\ref{nkern2}) (see the second reference
in \cite{nak63,nak69,nak88} for a discussion in the Wick-Cutkosky model).

In Ref. \cite{KW},  where a covariant framework  was adopted for
performing  the needed
analytic integrations, 
 the uniqueness theorem for the Nakanishi weight function 
\cite{nak71} was applied  directly to the BSE, obtaining an 
eigenequation like (\ref{newbound2})  and a kernel quite involved.
However,
the
 kernel shown in Appendix  C of Ref. \cite{KW}, is more general than the one shown in 
 Eqs. (\ref{newbound2}) and (\ref{nkern1}),
since a renormalized (at one loop) propagator and a sum of exchanged scalars
have been considered. Fortunately,  the presented numerical results  
were evaluated in
ladder approximation, as for the calculations  shown in Ref. \cite{carbonell1}
where the ladder approximation of Eq. (\ref{vbou}) was adopted. 
This motivated our investigation to only the ladder approximation, for the time
being. Indeed, the actual evaluation of the  ladder kernel shown in Eq. (\ref{nkern1}), allows one
to appreciate,  the well-known attractive feature of the 
LF framework to make less cumbersome the analytic integration, since the 
complexity of the calculation profitably distributes among two variables: 
$k^+$ and $k^-$  (see, e.g.,  Ref.
\cite{Sawicki} for a simple discussion of the box diagram).
 Finally, it is important to emphasize that Eq. (\ref{newbound2})  is an eigenequation, 
 with eigenvalue $1/\alpha$, once the mass $M$ of the interacting system 
 is assigned. Such a simple structure is a direct consequence of the linear
 dependence upon $\alpha$ of the kernel $i {\cal K}$, in ladder approximation.
  Differently,  Eq. (\ref{ptirl}) is a generalized eigenequation (cf Ref. \cite{carbonell1}).

As to the $\gamma$ dependence, we have already noted that for physical reason
(see Eqs. (\ref{valbt}) and (\ref{valbt1})), 
$g_b(z,\gamma;\kappa^2)$ must decrease for large values of $\gamma$.  Moreover,
one can check in ladder approximation that such a property is valid, since  a constant $g^{(Ld)}_b(z,\gamma;\kappa^2)$  for
$\gamma \to \infty$ leads to a different behavior for the lhs and rhs of Eq. 
(\ref{newbound2}). This can be seen by taking into account that in
 Eq. (\ref{hprime}) the difference between the second term and the third  one
becomes vanishing  for large $\gamma''$, and one remains with a $1/\gamma''$
fall-off on the rhs, namely the first term in Eq. (\ref{newbound2}), in contrast with the assumed constant behavior of 
$g^{(Ld)}_b(z,\gamma'';\kappa^2)$.

\section{LF-momentum Distributions}
\label{slf}
 It is attractive  to perform numerical comparisons that in perspective could be 
useful for an experimental investigation of actual
interacting systems. In view of this,  it  is very interesting to consider (see
the next Section for the actual calculations)   the  LF distributions of the 
valence component 
(cf  Eqs.
(\ref{val1}) and  (\ref{val2})). As shown below,   those distributions can be evaluated
through $g_b(z,\gamma;\kappa^2)$. Moreover, 
 the normalization of the valence component, once the BS amplitude itself 
 is properly normalized (see Appendix \ref{snorm} for
a short review of this issue and Refs. \cite{nak63,nak69,nak88,lurie} for details), yields the probability to find the valence 
contribution in the Fock expansion of the interacting two-scalar
state (see, e.g., \cite{Brodrev,dae}), viz
\be
 P_{val}= {1 \over (2 \pi)^3}~ \int_{0}^1 {d\xi \over 2~\xi(1-\xi)}~ \int 
 d{\bf k}_\perp~
 \psi^2_{n=2/p}(\xi,k_\perp)=\nonu =
 {1 \over (16 \pi)^2}~ \int_{-1}^1 dz~(1 -z^2)~ \int_0^{\infty} 
 d\gamma~ \left[\int_0^{\infty}d\gamma'~\frac{
g_b(\gamma',z;\kappa^2)}
{[\gamma'+\gamma +z^2 m^2+(1 -z^2)\kappa^2]^2}\right]^2
 \label{pval}\ee
 where Eq. (\ref{val1}) has been inserted in the last step, $\xi= (1-z)/2$ and 
 $ d{\bf k}_\perp ~= d\phi~ d\gamma/2$.  It should be reminded that
 $P_{val}\equiv N_2$, that is given in Eq. (\ref{N2}).

As is well known, one can defines the probability distribution to find a 
constituent with LF longitudinal
fraction $\xi=p^+_i/P^+$  in the valence state, as follows 
\be
\phi(\xi)= {1 \over (2 \pi)^3}~  {1\over 2~\xi(1-\xi)}~ \int 
 d{\bf k}_\perp~
 \psi^2_{n=2/p}(\xi,k_\perp)=
 \nonu=~2~
  {(1-z^2) \over (16\pi)^2}~ \int_0^{\infty} 
 d\gamma~\left [ \int_0^{\infty}d\gamma'~{
g_b(\gamma',z;\kappa^2)\over
[\gamma'+\gamma +z^2 m^2+(1 -z^2)\kappa^2]^2}\right]^2  
 \label{phixi}\ee
 with the obvious normalization: $
 \int_0^1 d\xi~\phi(\xi)=P_{val}
 $.
Furthermore,  one can consider  the probability distribution in
 $\gamma=|{\bf k}_\perp|^2$, i.e.
 \be
 {\cal P}(\gamma)= {1 \over 2(2 \pi)^3}~  \int_0^1 {d\xi\over 2~\xi(1-\xi)}~  
 \int_0^{2\pi} d\phi~
 \psi^2_{n=2/p}(\xi,k_\perp)=
 \nonu =
 {1 \over (16 \pi)^2}~  \int_{-1}^1 dz~ (1-z^2) 
 \left[ \int_0^{\infty}d\gamma'~
{g_b(\gamma',z;\kappa^2)
\over [\gamma'+\gamma +z^2 m^2+(1 -z^2)\kappa^2]^2}\right]^2
\label{probgam}\ee
with the normalization
 $
 \int_0^\infty d\gamma~{\cal P}(\gamma)=P_{val}
 $.

 Two final remarks are in order. Firstly, 
let us remind that for $\mu \to 0$ and $n=2$ in Eq. (\ref{naka1}), the Nakanishi amplitude factorizes as $g_b(\gamma',z;\kappa^2)
\to \delta(\gamma')~ f(z;\kappa^2)$ (see, e.g., \cite{dae}), and therefore in
the Wick-Cutkosky model one gets
\be
\psi^{WiC}_{n=2/p}(\xi,k_\perp)~
 \propto~{f(z;\kappa^2)
\over [\gamma +z^2 m^2+(1 -z^2)\kappa^2]^2}  
\ee

Secondly, we would emphasize that the valence wave function behaves as
expected (see \cite{Brodrev}) for large values of $k^2_\perp=\gamma$, once we choose $n=2$. As a matter of fact, the
Nakanishi weight function drops out for increasing $\gamma'$, and one has  for $\gamma \to \infty$
\be
\psi_{n=2/p}(\xi,k_\perp)=
\nonu =~{(1-z^2)\over 4\sqrt{2}}
~ \int_0^{\infty}d\gamma'~
{g_b(\gamma',z;\kappa^2)
\over [\gamma'+\gamma +z^2 m^2+(1 -z^2)\kappa^2]^2}  \to ~
{C(z)\over \gamma^2 }
\ee
with a $\gamma$-tail independent upon the mass of the exchanged scalar.

In  the next Section,  the numerical results of the LF
distributions, obtained in ladder approximation, 
are presented. We can anticipate that such  LF distributions, evaluated 
by using   the  solutions of Eqs. (\ref{ptirl}) and
(\ref{newbound2}) for  a given mass of the exchanged meson and binding energy,
overlap, though the numerical Nakanishi weight functions, 
$g^{(Ld)}_b(\gamma',z;\kappa^2) $,  show few-percent differences 
 for  low values of
$\gamma$, as discussed in what follows.

\section{Numerical Comparisons}
\label{ris}
In order to implement  the quantitative studies of  
the Nakanishi weight function  for the S-wave BS amplitude 
of a two-scalar system, with a massive scalar exchange, we have adopted a
proper  basis. This basis  allows us to expand 
 the non singular weight function by  taking into account the 
  features of 
$g^{(Ld)}_b(\gamma,z;\kappa^2)$ discussed in Sects. \ref{sbslf} and \ref{slad};
namely
 i) the symmetry 
with respect to $z$, ii) the constraint $g^{(Ld)}_b(\gamma,z=\pm 1;\kappa^2)=0$   and iii)
 the fall-off in $\gamma$. In particular,  Gegenbauer
polynomials with proper indexes have been chosen for describing the $z$ dependence, while the Laguerre polynomials have 
been adopted for 
the  $\gamma$-dependence. In short, we have expanded the Nakanishi weight
function as follows
\be g^{(Ld)}_b(\gamma,z;\kappa^2) = \sum_{\ell=0}^{N_z} \sum_{j=0}^{N_g}~
A_{\ell j} ~G_\ell(z) ~{\cal L}_j(\gamma)
\label{bas1}\ee
where i) the functions $G_\ell(z)$ are given in terms of even Gegenbauer
 polynomials, $ C^{(5/2)}_{2\ell}(z)$  by 
\be
G_\ell(z)= 4~(1-z^2) ~\Gamma(5/2)~\sqrt{{(2\ell+5/2) ~(2\ell)! \over \pi
 \Gamma(2\ell+5)}}~ C^{(5/2)}_{2\ell}(z)
\label{bas2}\ee
and ii)
the  functions ${\cal L}_j(\gamma)$ are expressed in terms of the Laguerre polynomials,  $
L_{j}(a\gamma)$, by
\be
{\cal L}_j(\gamma)= \sqrt{a}~ L_{j}(a\gamma) ~
e^{-a\gamma/2}~~~~~~~~~~.\label{bas3}\ee
The following orthonormality conditions are fulfilled
\be
\int^1_{-1}dz~G_\ell(z)~G_n(z)=\delta_{\ell n} ~~~~~~~~,
\nonu
  \int_0^\infty~ d\gamma~{\cal L}_j(\gamma)~{\cal L}_\ell(\gamma)=
a \int_0^\infty~ d\gamma~ e^{-a\gamma}~L_{j}(a\gamma) ~L_{\ell}(a\gamma)=
~\delta_{j\ell}
\ee
In order to speed up the convergence, in the actual calculations the parameter 
$a=6.0$ 
has been adopted, and   the variable $\gamma$ has been rescaled according to $\gamma 
\to 2 \gamma/a_0$  with
$a_0=12$.  It is worth noting that the two parameters $a$ and $a_0$ control,
 loosely speaking, the range of
relevance of the Laguerre polynomials and the structure of the kernel, 
respectively.
Finally, the integration over the variable $z$ has been performed by using 
 a Gauss-Legendre quadrature rule,
 while the
Gauss-Laguerre quadrature  has been adopted for  
the variable $\gamma$.

\subsection{Eigenvalues and Eigenvectors}
We have first solved Eq. (\ref{ptirl}), i.e. the one proposed in Ref.
\cite{carbonell1}, but  using our basis instead of the spline basis adopted there.
With the spline basis, for both $z$ and $\gamma$, some instabilities appear
and in \cite{carbonell1} a small parameter was introduced to achieve stable
results (see also below). Our basis allows us to overcome such a problem, since
it contains the above mentioned general features of 
$g^{(Ld)}_b(\gamma,z=\pm 1;\kappa^2)$. This first step was necessary to gain  confidence in our
 basis, through the comparison with the results in \cite{carbonell1} (see what follows).
As a  second step, we evaluated  eigenvalues and
 eigenvectors of Eq. (\ref{newbound2}), which was deduced by invoking the uniqueness theorem.
 As for this equation, it should be pointed
out that a completely different numerical method was chosen in \cite{KW}.
In particular, it was applied an iterative procedure, suggested by the structure 
of the ladder kernel obtained in \cite{KW}.

 In the following Tables a detailed comparison between our results and the ones  obtained 
in  Refs. \cite{carbonell1} and \cite{KW} is presented. Let us remind that in
\cite{KW}, though the proposed ladder kernel contains dressed propagators and a
sum of exchanged meson, the numerical evaluations were performed without such
extras, and therefore their results can be directly compared to ours and the
ones in \cite{carbonell1}, with only the caveat of a different definition of the
coupling constant $\alpha$.
 As already pointed out in Refs. \cite{KW} and \cite{carbonell1}, the
kernel contains a  
 highly non
linear dependence upon the mass $M$ of the interacting system, but a linear dependence upon 
the coupling constant $\alpha$, given the  adopted ladder
approximation. Therefore, it is customary i) first to choose  a value for the binding 
energy  in the interval $$0~\leq ~{B \over m}~= ~
2 -{M\over m} ~\leq ~2~~~~~~~~,$$ and ii) then to look for the minimal value of the
 coupling constant that allows such a binding energy. 
  A comment on the range of the usually-chosen interval is in order.
 As is well-known (see Ref. \cite{gbaym}), all the $\phi^3$ models do not show
 any ground state, nonetheless they are widely adopted for illustrative purposes
 and for gaining insights into the effectiveness of theoretical tools. Here, we also adhere to this general attitude
  (see Ref. \cite{gross} for some details on how
 and to what
 extent it is possible to
  reconcile the general features of the $\phi^3$ and the actual calculations). After introducing a basis,
 it should be  noticed that in the case 
 of Eq. (\ref{ptirl}), one has
 a generalized eigenvalue problem (cf Ref. \cite{carbonell1}), that in a
 symbolic form reads
 \be {1\over \alpha}~ B(M) ~g^{(Ld)} = A^{(Ld)}(M)~g{(Ld)}\label{symb}\ee
 while for the Eq. (\ref{newbound2}) one has a genuine eigenvalue problem, viz
 \be {1\over \alpha}~ g = D^{(Ld)}(M)~g{(Ld)}\ee
 The possibility to reduce the first problem to the second one relies on the existence of the inverse
 of the integral operator $B(M)$, and the numerical feasibility of such 
 inversion with enough accuracy. 
 In particular, in Ref. \cite{carbonell1}, where the spline basis was adopted, 
 a small parameter was added to the matrix $B(M)$
 in order to achieve a good stability. We have investigated if adopting our basis, 
 that includes the expected fall-off of the weight function
 for large values of $\gamma$, one has to similarly introduce a small parameter.
  Fortunately, with our basis,
 the small quantity to be added to the diagonal terms of $A^{(Ld)}(M)$ is
 $\epsilon=10^{-9}$ (the largest number of Gaussian points was 80 for each variable in 
 $g^{(Ld)}$). As for Eq. (\ref{newbound2}), one has been able to get rid of the numerical 
 inversion
 of the matrix, since, de facto, it has been mathematically performed.
 Finally, it is important noticing that, 
for  both equations,   the involved matrices are real but not symmetric, and
therefore  
pairs of complex eigenvalues can appear.

 In order to achieve a very good convergence for both eigenvalues and
 eigenvectors (in particular for Eq. (\ref{newbound2})), the numerical studies
  with 
 the basis in Eqs. (\ref{bas1}),
 (\ref{bas2}) and (\ref{bas3}) has been extended up to 
 $N_z=18$ and $N_g=32$, for all the values of $B/m$, except for $B/m=0.01$ where
 we extend $N_g$ up to $48$. Indeed, for $B/m \geq 0.1$
  a nice stability  of the eigenvalues can be reached already for 
  $N_z=8$ and $N_g> 24$. In general, the stability of the eigenvalues settles
  well before than  the convergence of the eigenvectors.  
\begin{table}
\caption
{ Values of $\alpha=g^2/(16 \pi m^2)$, obtained by solving the eigenequations
  (\ref{ptirl}) and 
(\ref{newbound2})
 (i.e. the eigenequation with the
application of the uniqueness theorem). Results correspond to 
 $\mu/m=0.15,~0.50$, varying the  binding energies, $B/m$. 
 The second column contains
the results obtained in Ref. \cite{carbonell1} by using the spline basis 
and Eq. (\ref{ptirl}); 
the third column shows  our results obtained from Eq. (\ref{ptirl}) by using our
 basis (Eqs. (\ref{bas1}), (\ref{bas2}) and  (\ref{bas3})), with $N_z=18$,
 $N_g=32$  and 
 $a=6$  in Eq.
 (\ref{bas3}) ); the 
fourth column  contains our results  obtained from the eigenequation 
(\ref{newbound2}) and our basis.
 ($^*$) For $\mu/m=0.15$ and $B/m=0.01$, 
   the stability of the coupling constant ($\alpha<1$) is reached
 for $N_g\ge 46$. }

\parbox {6.5cm}{
\begin{tabular}{c} 
$ \mu/m=0.15  $  \end{tabular}
\begin{tabular}{|c|c|c||c|}
\hline
 B/m  & $\alpha$ \cite{carbonell1}  & $\alpha$ Eq. (\ref{ptirl}) & $\alpha$ Eq. (\ref{newbound2})\\
 \hline
  0.01 & 0.5716  &  0.5716&  0.5716($^*$)\\ \hline
  0.10 & 1.437   &  1.437 &  1.437 \\ \hline
  0.20 & 2.100   &  2.099 &  2.099 \\ \hline
  0.50 & 3.611   &  3.610 &  3.611 \\ \hline
  1.00 & 5.315   &  5.313 &  5.314 \\ \hline
\end{tabular}} 
 $~~~~~~$
\parbox {6.5cm}{\begin{tabular}{c} 
$ \mu/m=0.50  $  \end{tabular}
\begin{tabular}{|c|c|c||c|}
\hline
 B/m   &  $\alpha$\cite{carbonell1}  & $\alpha$ Eq. (\ref{ptirl}) & $\alpha$ Eq. (\ref{newbound2})\\
 \hline
  0.01 &  1.440 & 1.440& 1.440\\ \hline
  0.10 &  2.498 & 2.498& 2.498\\ \hline
  0.20 &  3.251 & 3.251& 3.251\\\hline
  0.50 &  4.901 & 4.901& 4.901\\ \hline
  1.00 &  6.712 & 6.711& 6.711\\ \hline
\end{tabular} } 
\label{tab1}
\end{table}
\begin{table}
\caption
{ Values 
of $\alpha=g^2/(16 \pi m^2)$,  obtained by solving the eigenequations
 (\ref{newbound2}) (i.e. with the
application of the uniqueness theorem) and  (\ref{ptirl}). Results correspond to  
  $\mu/m=0.50$, 
varying the binding energies, $B/m$. The second column shows
the values obtained in Ref. \cite{KW}, where the uniqueness theorem was
exploited and  an iterative method was adopted; 
the third column corresponds to the solution of Eq. (\ref{newbound2}) 
by using our basis (cf Eqs. (\ref{bas1}),
(\ref{bas2}) and  (\ref{bas3}) ); the
fourth column contains our results from Eq.  (\ref{ptirl}).}
\parbox {6.5cm}{ 
\begin{tabular}{c} 
$\mu/m=0.50 $  \end{tabular}
\begin{tabular}{|c|c|c||c|}
\hline
  B/m   &  $\alpha$\cite{KW} & $\alpha$ Eq. (\ref{newbound2}) & $\alpha$ Eq. (\ref{ptirl}) \\ \hline
  0.002 & 1.211 & 1.216   & 1.216    \\ \hline  
  0.02  & 1.624 & 1.623   & 1.623     \\ \hline 
  0.20  & 3.252 & 3.251   & 3.251 	  \\ \hline
  0.40  & 4.416 & 4.415   & 4.416        \\ \hline
  0.80  & 6.096 & 6.094   & 6.094     \\ \hline
  1.20  & 7.206 & 7.204   & 7.204     \\ \hline
  1.60  & 7.850 & 7.849   & 7.849     \\ \hline
  2.00  & 8.062 & 8.061   & 8.061     \\ \hline
\end{tabular}}  
\label{tab2}
\end{table}
 
   In Table \ref{tab1}, the results for the coupling constant
   $\alpha$, corresponding to   
Eq. (\ref{ptirl}) and Eq. (\ref{newbound2}), for 
  $\mu/m= 0.15,~ 0.50$ 
  and a set of binding energies, $B/m$, are shown. 
In particular, in the second column,  the results 
obtained in  \cite{carbonell1} 
by using the spline basis are reported, while our results corresponding to
both Eq.
(\ref{ptirl}) and Eq.
(\ref{newbound2}) are presented in the third column and the fourth one, 
respectively. It is important to note that for 
 $B/m=0.01$ and $\mu/m=0.15$,  the stability of the eigenvalue 
 obtained through Eq.
(\ref{newbound2}) is   reached with $N_g \ge  46$, when $a=6$ is
chosen, while $N_g=28$ is enough when $a=12$ is adopted (with this value for $a$, the 
convergence
of the eigenvectors is not satisfactory for $N_g=28$).

In Table  \ref{tab2}, it is presented the comparison with the results from Ref. \cite{KW}, where
the uniqueness theorem 
was used. It should be pointed out  that in  Ref.
 \cite{KW} only the value $\mu/m=0.50$  was considered, and the coupling constant
 contained an extra factor $\pi$ with respect to the definition adopted in the
 present paper and in \cite{carbonell1}.   It is important to remind that
  the eigenvalues shown in 
  Ref. \cite{KW}
compared very favorably with the ones obtained in Ref. \cite{LM}, where the BS 
equation in ladder approximation
was solved in Euclidean space. 
 Moreover, one can find in Refs.
\cite{carbonell7,maris} more evaluations both 
within the LF Hamiltonian dynamics and in Euclidean space, that appear in nice
agreement with our calculations. 

Finally, it
  should be pointed that all the digits of our results presented in the Tables 
  are stable, and the 
   numerical uncertainties affect only the digit beyond the ones shown, at the
   level of a few units.
   
\begin{figure}[th]
\includegraphics[width=10.cm]{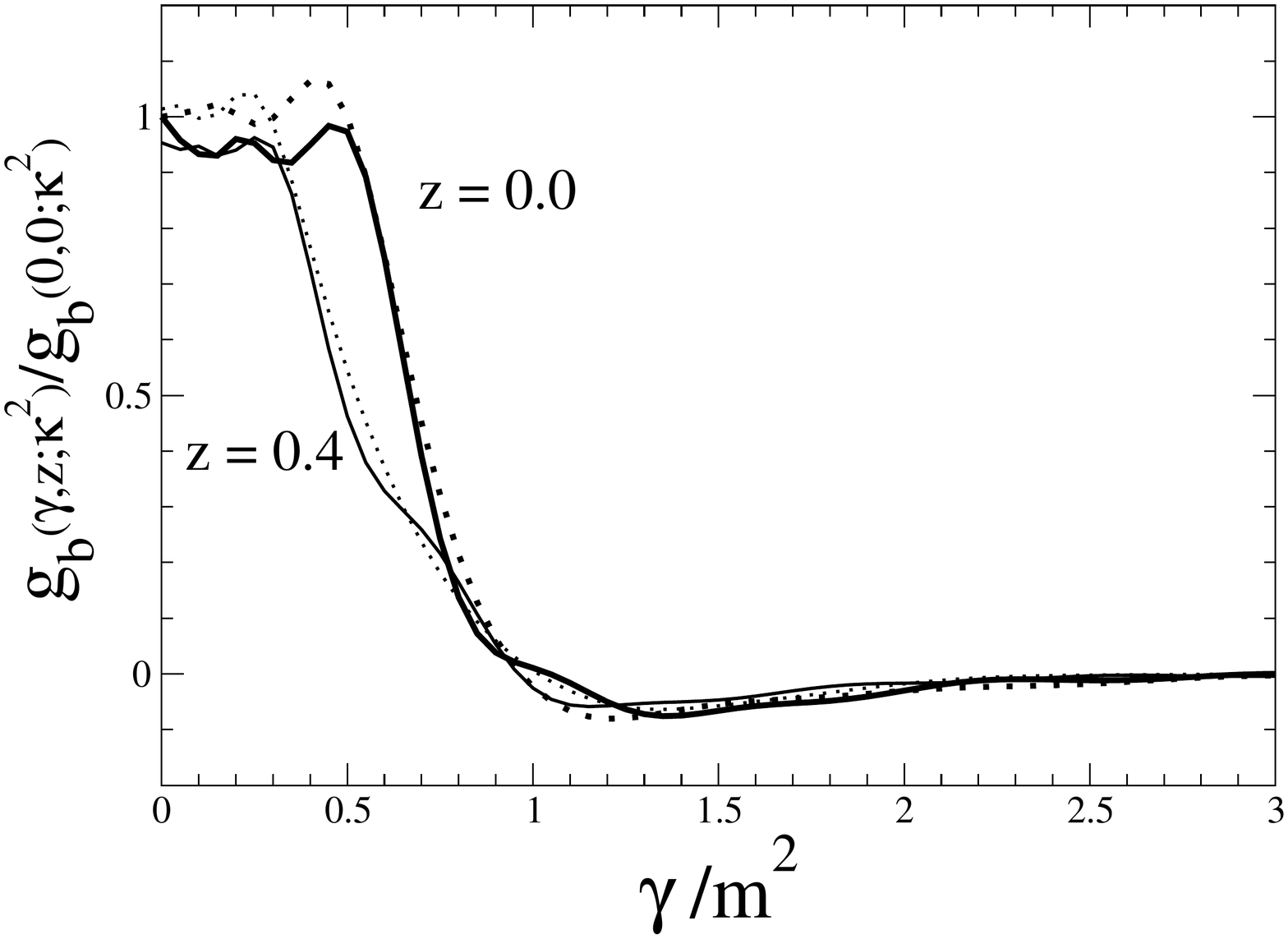}

\includegraphics[width=10.cm]{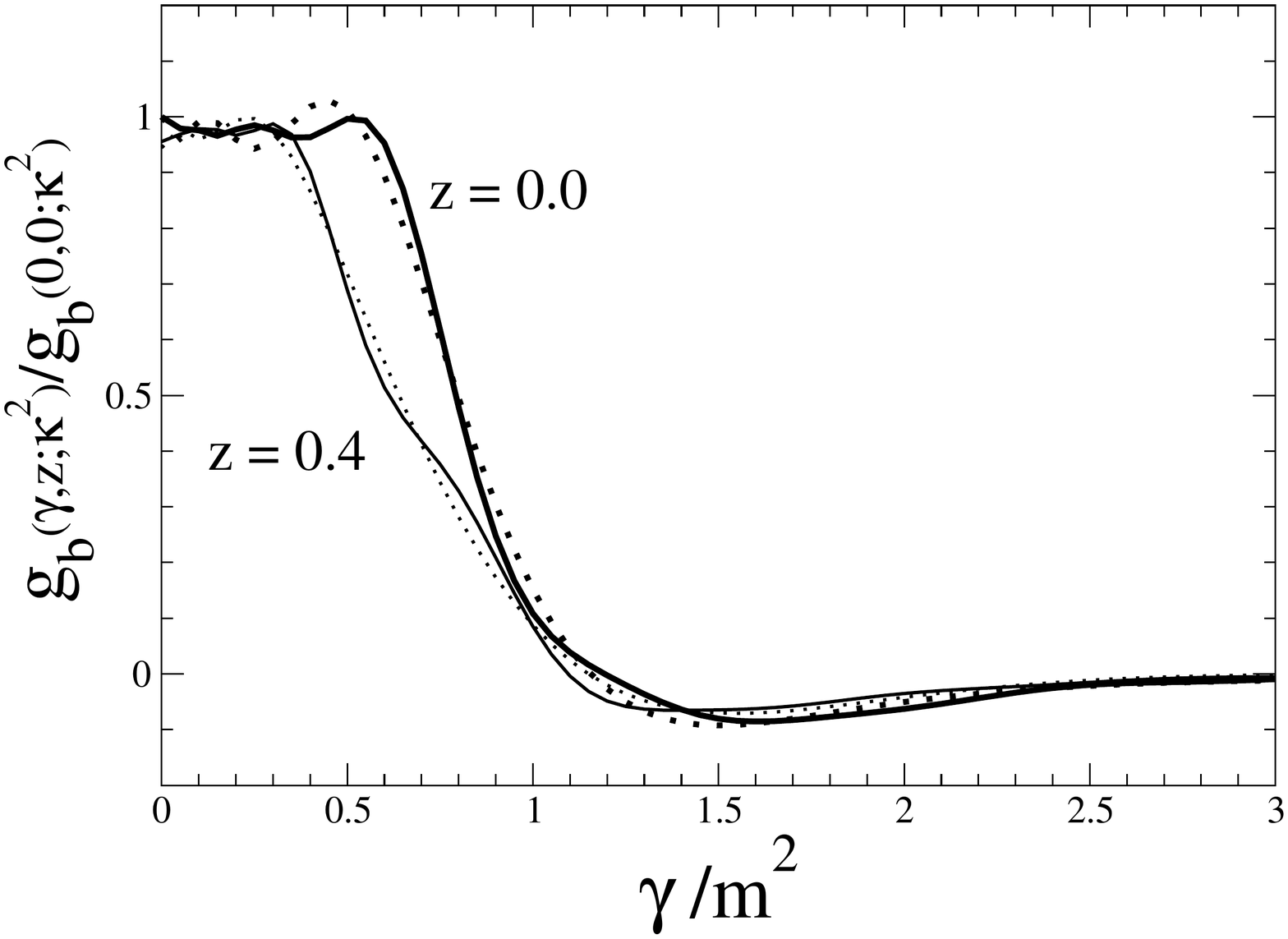}

\includegraphics[width=10.cm]{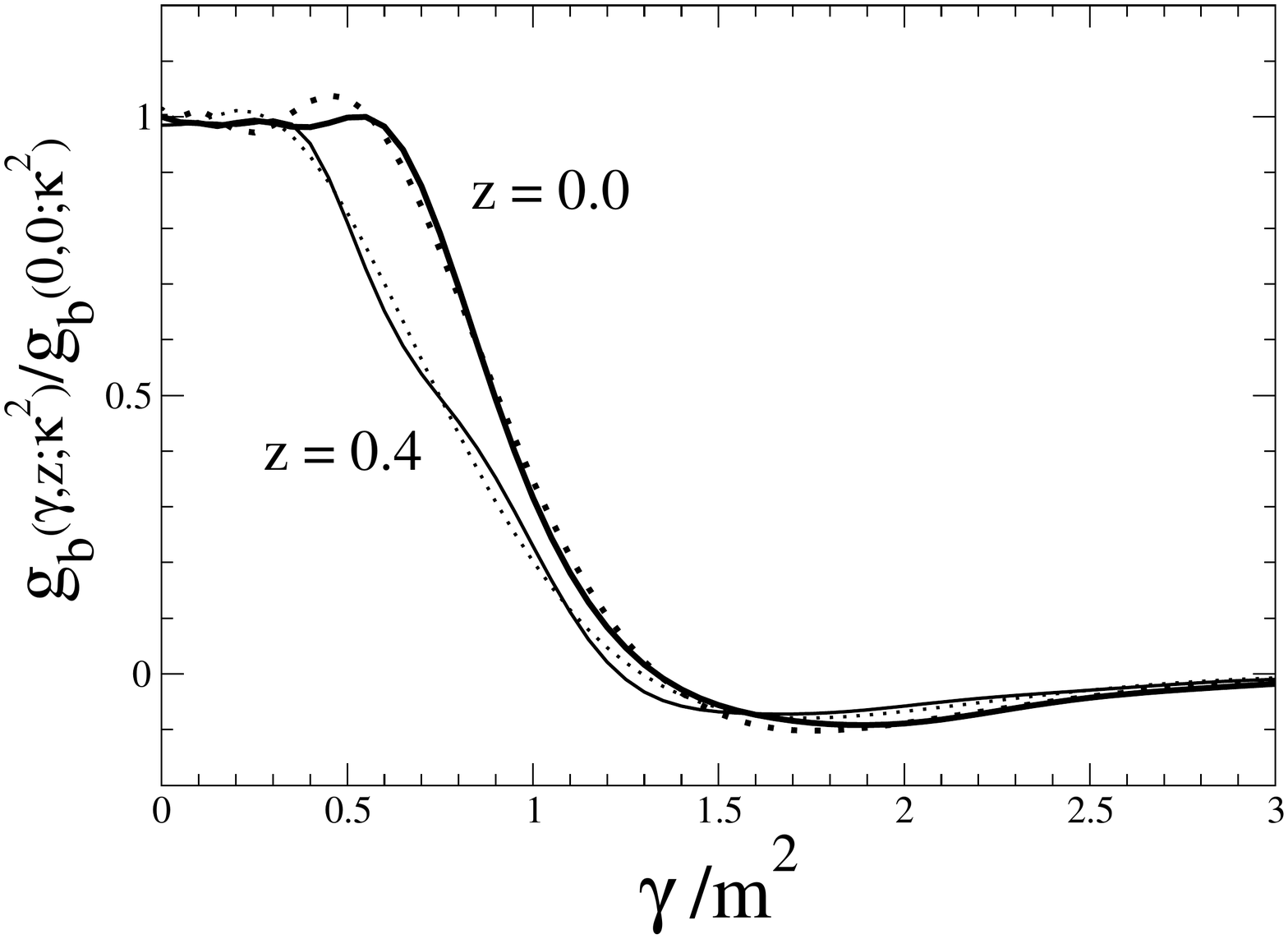}

\caption{The Nakanishi weight function $g_b^{(Ld)}(\gamma,z;\kappa^2)$ 
for $\mu/m=0.5$ and 
$B/m=0.2,~0.5,~1.0$ (from the top) vs $\gamma/m^2$ and two values of $z$. Thick lines refer to $z=0$ and
thin lines to $z=0.4$, as indicated by the inset. Solid lines: results from 
Eq. (\ref{newbound2}). Dotted lines: results from 
Eq. (\ref{ptirl}).}
\label{g_gam}
\end{figure}

\begin{figure}[th]
\includegraphics[width=8.cm]{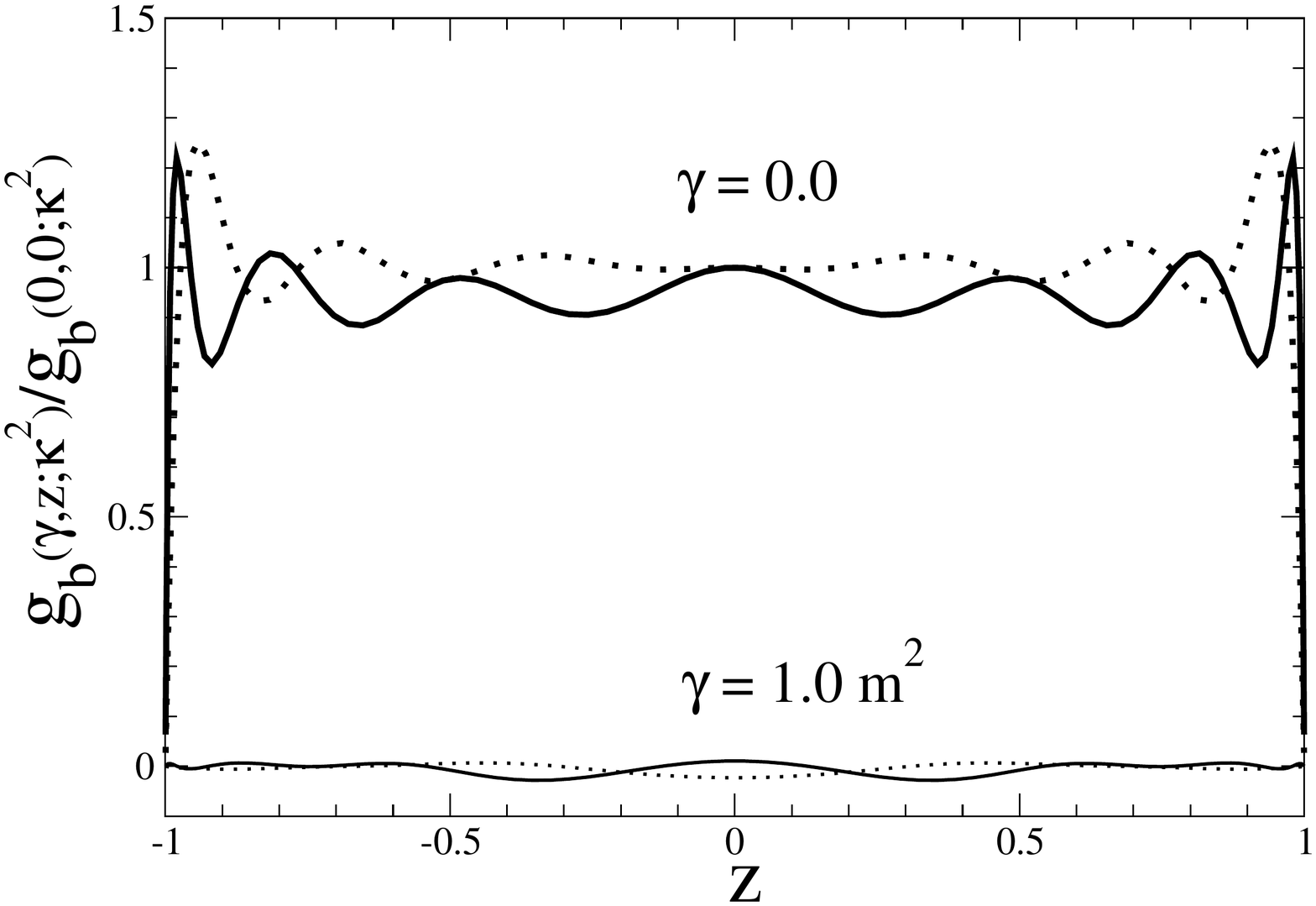} 
\includegraphics[width=8.cm]{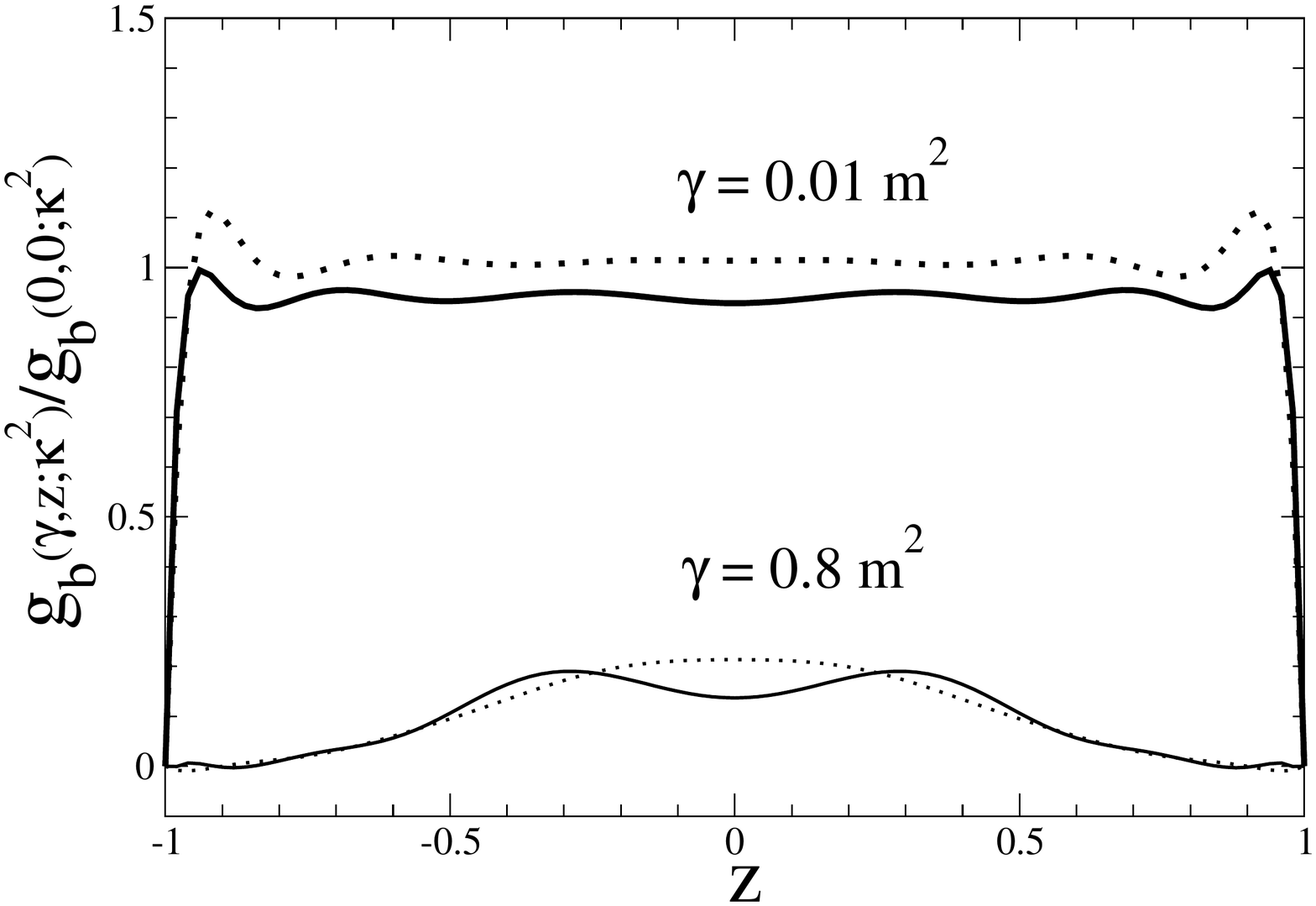}

\includegraphics[width=8.cm]{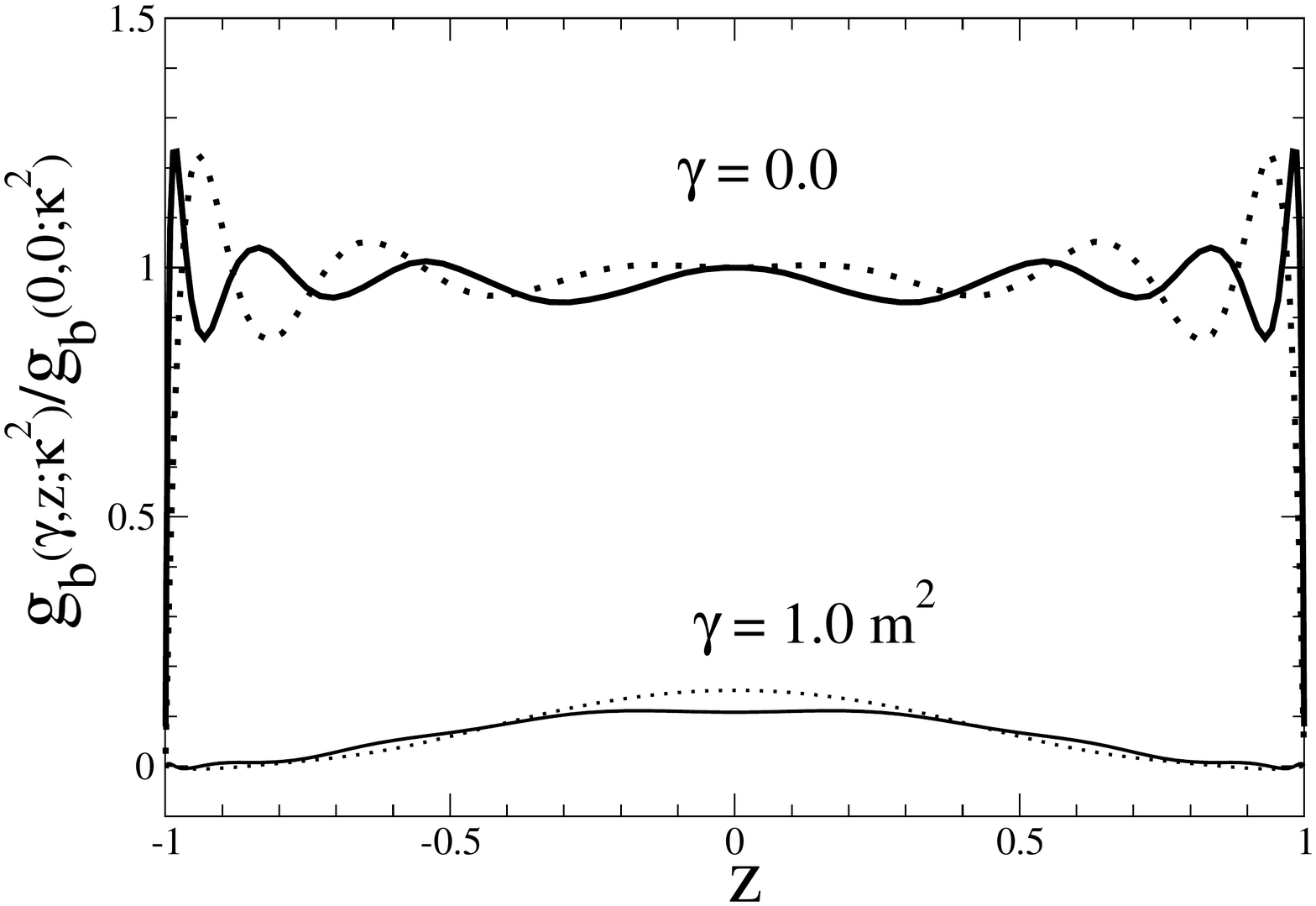}
\includegraphics[width=8.cm]{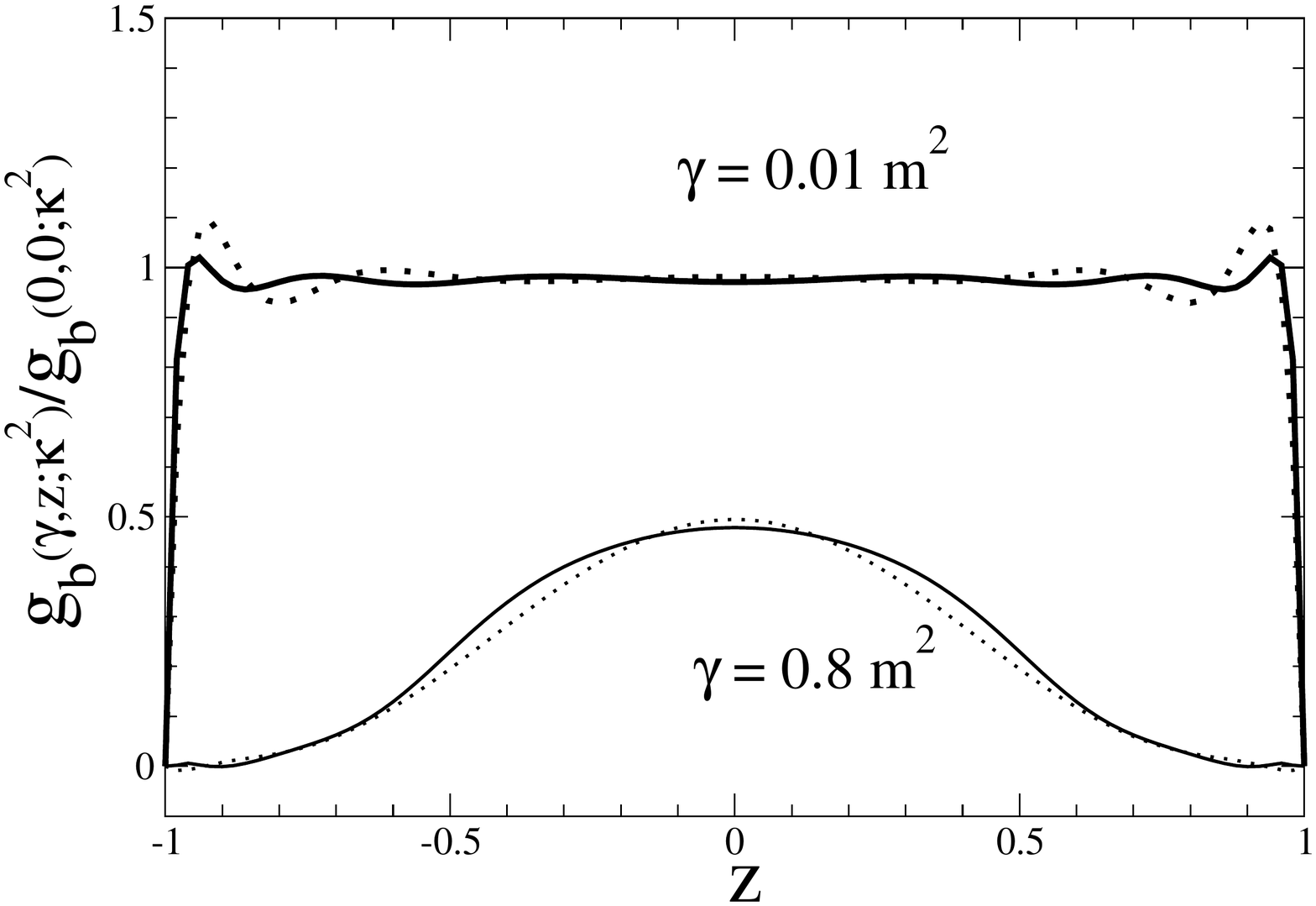}

\includegraphics[width=8.cm]{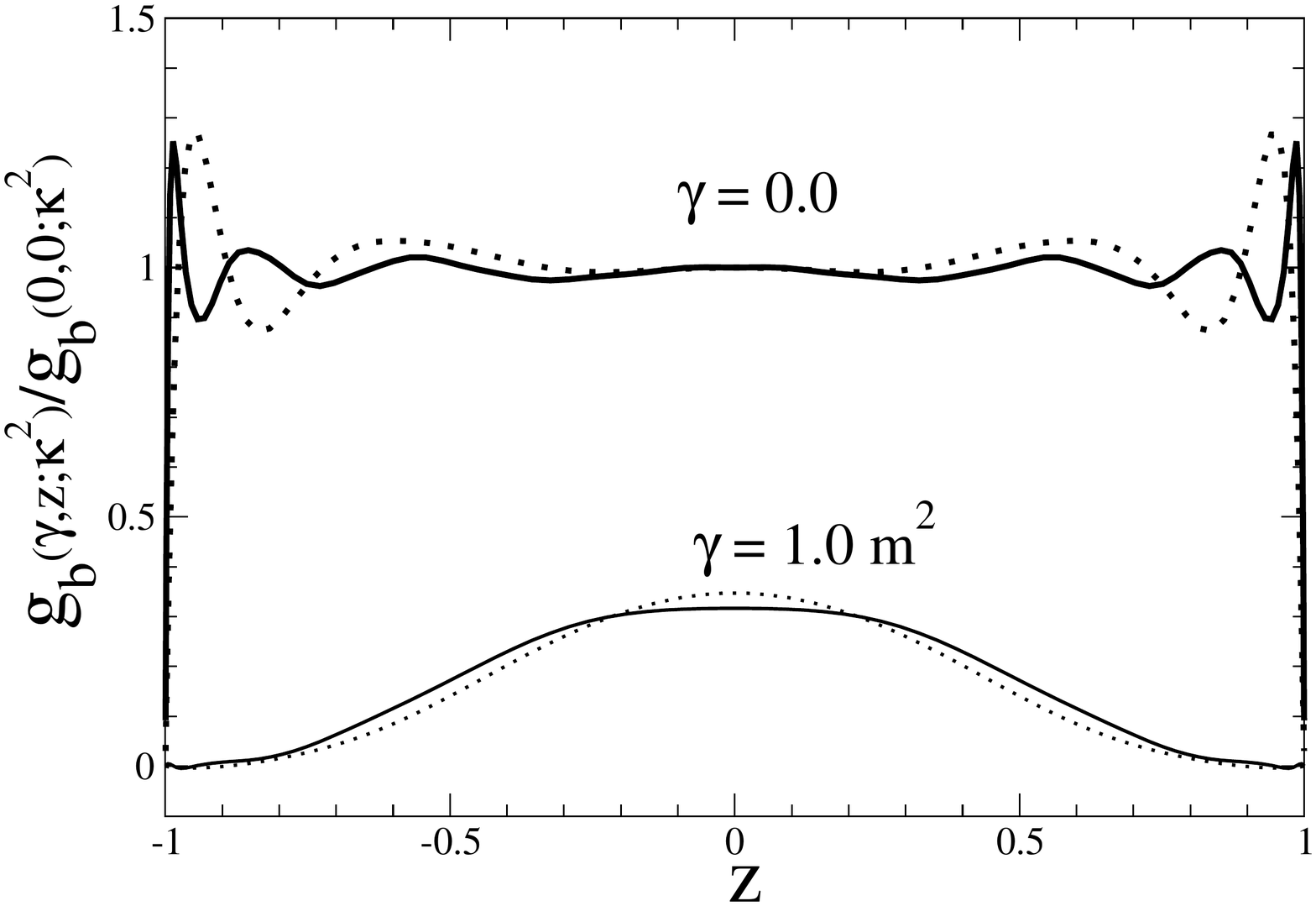}
\includegraphics[width=8.cm]{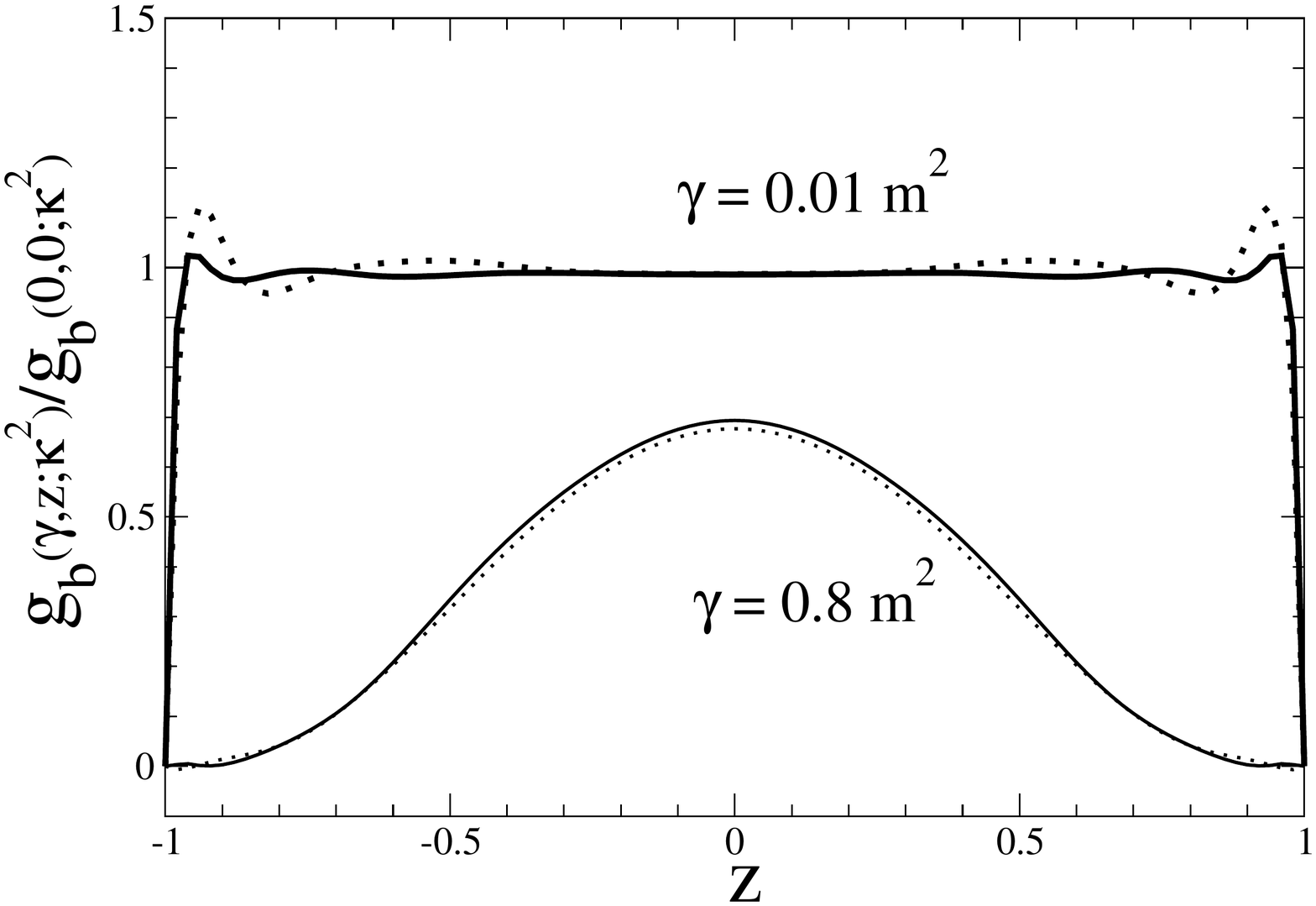}
\caption{The Nakanishi weight function $g_b^{(Ld)}(\gamma,z;\kappa^2)$ for $\mu/m=0.5$ and 
$B/m=0.2,~0.5,~1.0$ (from the top) vs $z$ and four values of $\gamma/m^2$. Thick lines refer to 
$\gamma=0$ and $\gamma=0.01~m^2$, while 
thin lines to $\gamma=0.8~m^2$ and  $\gamma=1~m^2$, as indicated by the inset. Solid lines: results from 
Eq. (\ref{newbound2}). Dotted lines: results from 
Eq. (\ref{ptirl}).}
\label{g_z}
\end{figure}

In Figs. (\ref{g_gam}) and    (\ref{g_z}) the comparison between 
 the  weight functions 
obtained from Eqs. (\ref{ptirl}) and 
(\ref{newbound}) is shown  for   $\mu/m=0.50$ and $B/m=0.2,~0.5, ~1.0$.     
Few-percent differences appear for small values of $\gamma$, and are bigger for
small values of the binding energy. In this case, the characteristic momentum associated with the weak-binding
 energy is much smaller than the mass scale of the 
system, and therefore to   appropriately describe the Nakanishi weight function one should use  a 
 larger basis which  accurately spans
 both the small and large momentum regions. This demands  more  numerical
  efforts, that can be postponed, since our present  aim is  to validate the Nakanishi approach  over 
  the   largest  range of dynamical regimes, which can be covered by the basis we have chosen (see, e.g., Table \ref{tab1} and $B/m=0.01$ and
  $\mu/m=0.15$).
  Notably, the above mentioned  differences do not have
any sizable effect on the LF distributions (see the next subsection).

As a further check, we evaluated the solution of  Eq. (\ref{ptirl}), corresponding to $\mu/m=0.50$
and $B/m=1.0$, by introducing a small parameter as in \cite{carbonell1}. In particular we adopted $\epsilon= 10^{-4}$ for comparing with the 
 weight function presented in Figs.  2 and 3 in Ref. \cite{carbonell1}, and we obtained the same results.
  It is worth noting that also by adopting the small parameter $\epsilon= 10^{-4}$, we did not find   any sizable effects on the LF
distributions. 
\subsection{Valence probability and LF distributions}
After determining  the expansion coefficients of the Nakanishi weight function, as given in  
 Eq. (\ref{bas1}), 
  and imposing the normalization condition on the BS amplitude, Eq. (\ref{normg}), 
 one can calculate the valence component
of the interacting system, Eq.(\ref{val1}).  Then, very interesting (in particular from the phenomenological 
point of view) quantities can be evaluated.
First of all, the valence probability, Eq. (\ref{pval}), can be obtained, 
The results are shown in Table
\ref{tab4} for $\mu/m=0.05$, $\mu/m=0.15$  and $\mu/m=0.5$.
Several values of $B/m$ have been chosen for covering the interval $0<B/m\leq 2$. 
It should be recalled that the asymptotic value $P_{val}=1$, reached 
 for $B/m \to 0$, is more and more closely approached for
 smaller and smaller values of $B/m$ (or equivalently smaller values of
 $\alpha$) when  $\mu/m$ decreases. Since in Table \ref{tab4} there are
 also the results for $\mu/m=0.05$, a by-product of these calculations is the
 following 
interesting remark.  For $B/m=2$ and  decreasing  $\mu/m$,  the values of $\alpha$ 
 show a  decreasing behavior 
 toward $\alpha=2\pi$, namely 
 the value of $\alpha$  obtained in  the Wick-Cutkosky case, 
 i.e. $\mu/m=0$ (cf  Ref. \cite{dae}). Correspondingly  the valence probability approaches the
 Wick-Cutkosky value   $P_{val} \sim 0.64$ \cite{dae}.

\begin{figure}
\includegraphics[width=9.cm]{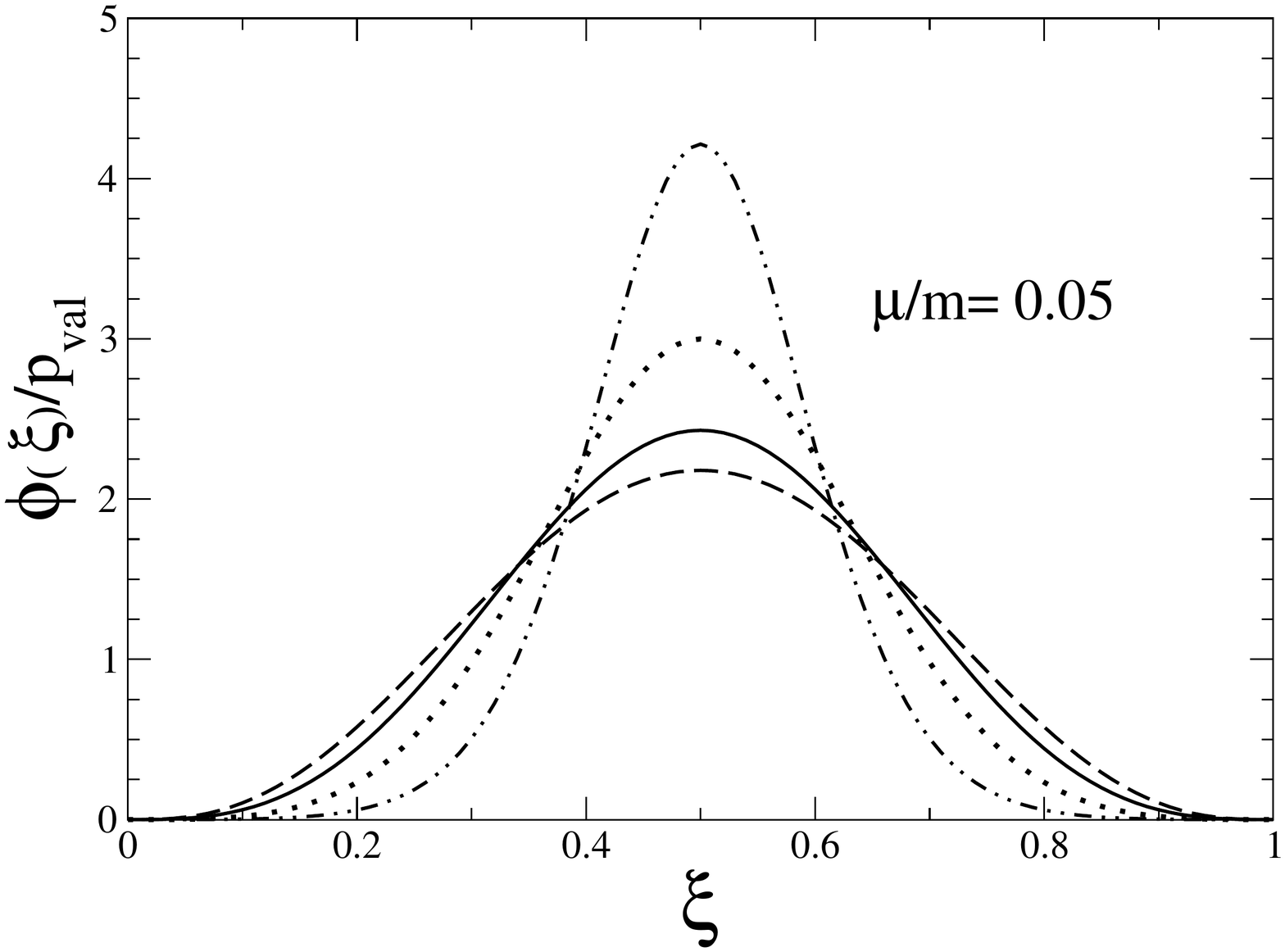}

\includegraphics[width=9.cm]{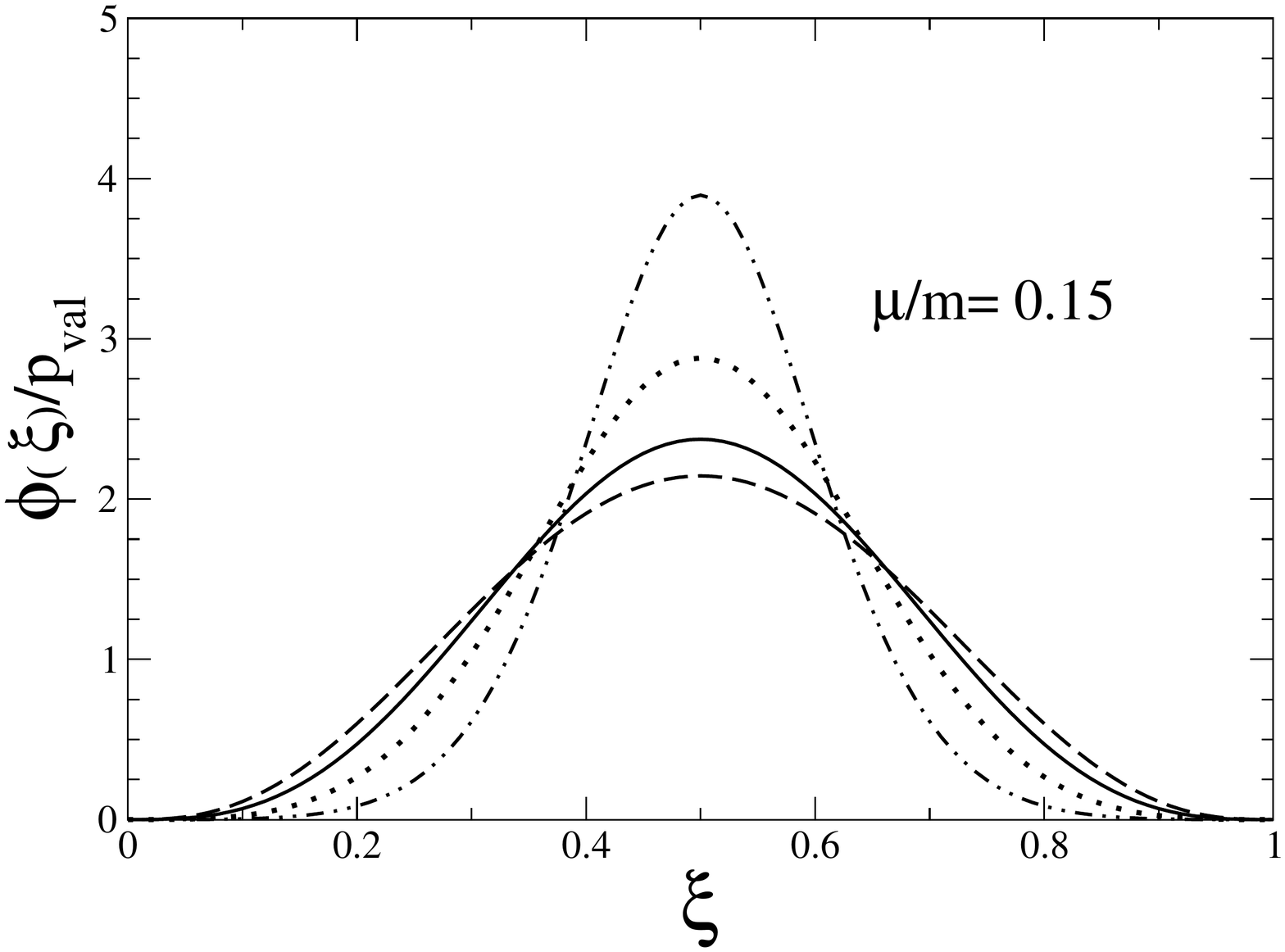}

\includegraphics[width=9.cm]{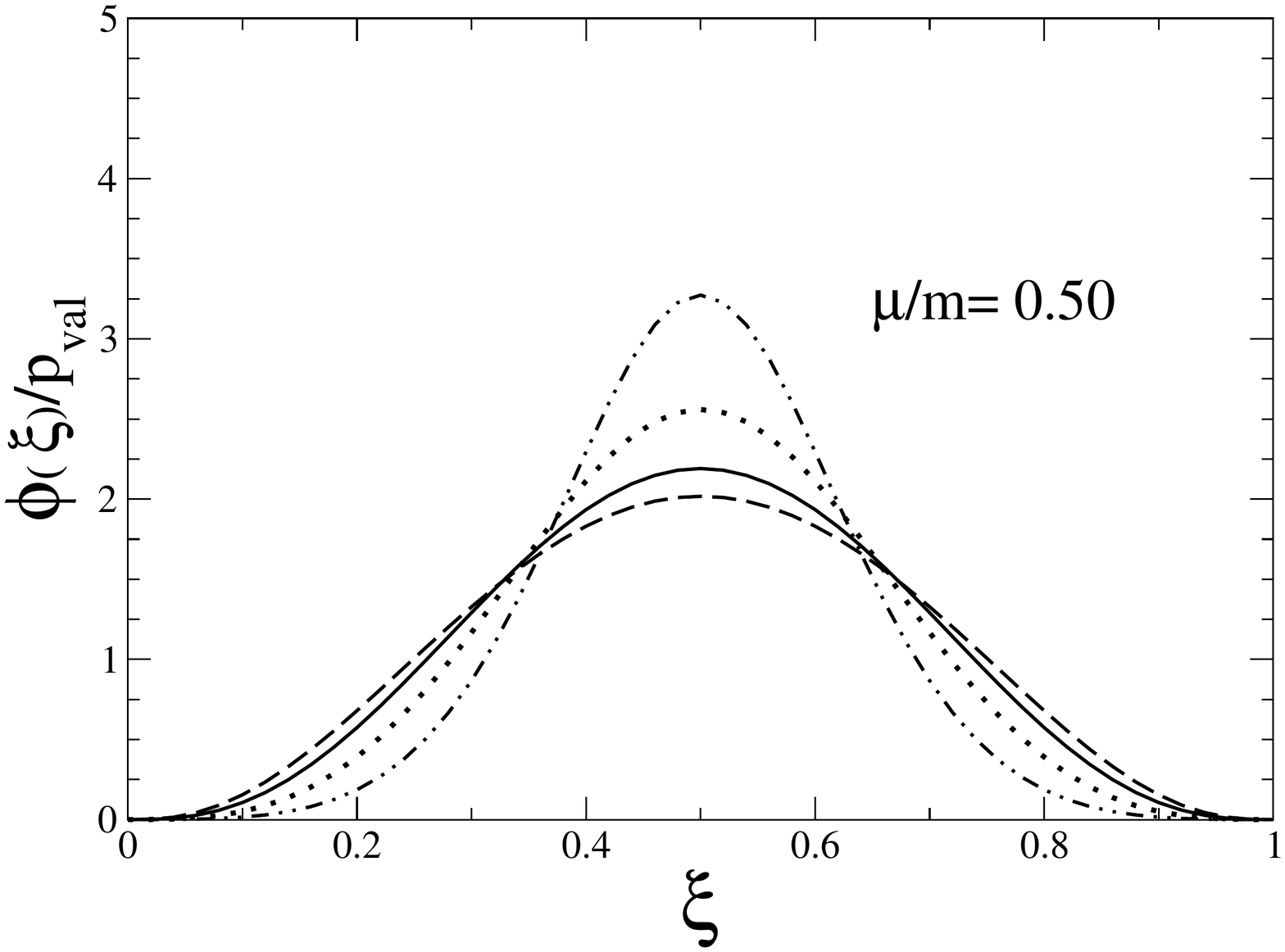}
\caption{The longitudinal LF-distribution $\phi(\xi)$ for the valence component, Eq. (\ref{phixi}), vs the longitudinal-momentum fraction 
$\xi$, for 
$\mu/m= 0.05,~0.15~0.50$. Dash-double-dotted line: $B/m=0.20$. Dotted line: $B/m=0.50$. 
Solid line: $B/m=1.0$. Dashed line: $B/m=2.0$. Recall that $\int_0^1 d\xi~\phi(\xi)=P_{val}$ (cf 
Table \ref{tab4}).}
\label{fixi}
\end{figure}

\begin{figure}
\includegraphics[width=9.cm]{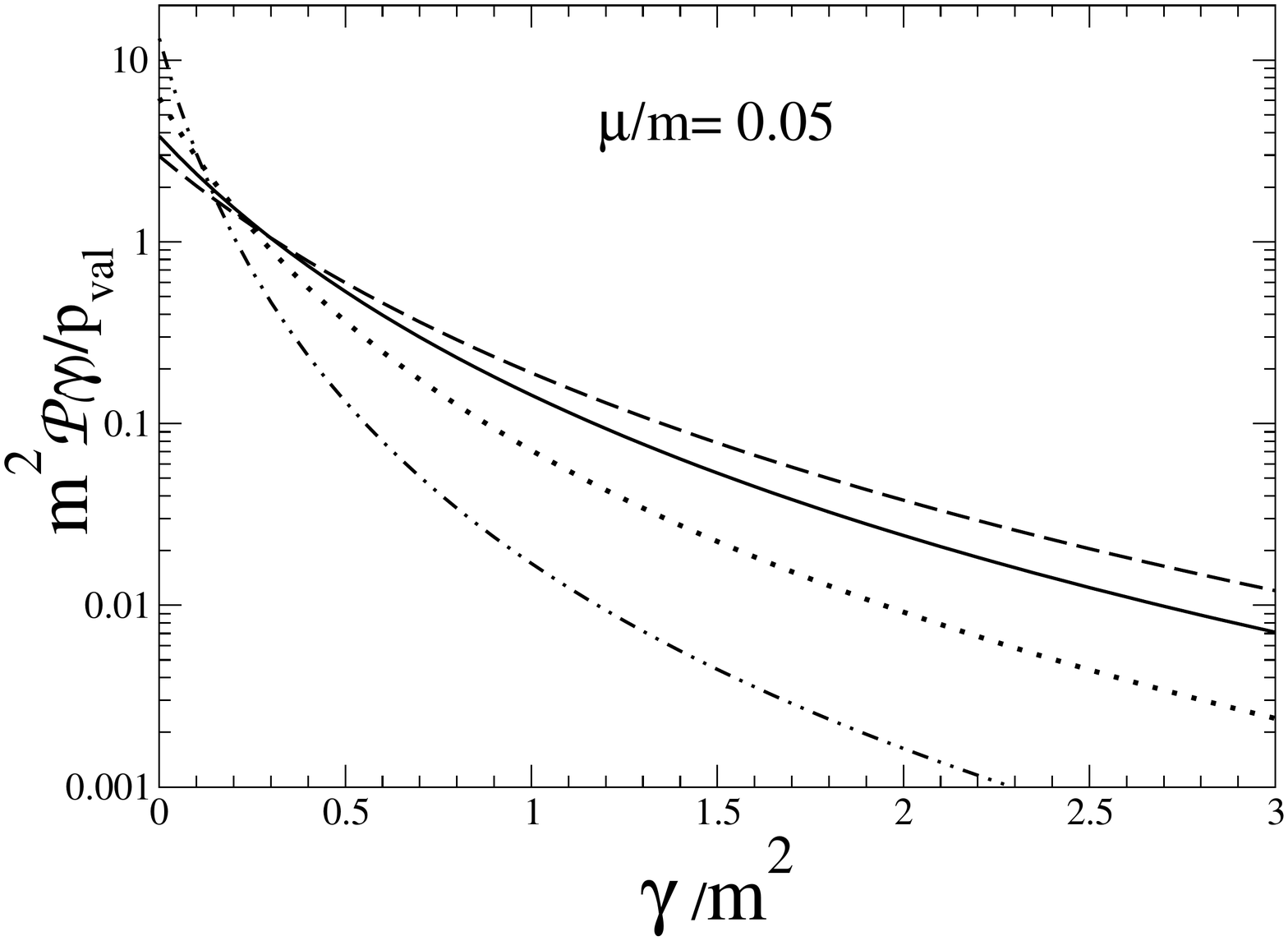}

\includegraphics[width=9.cm]{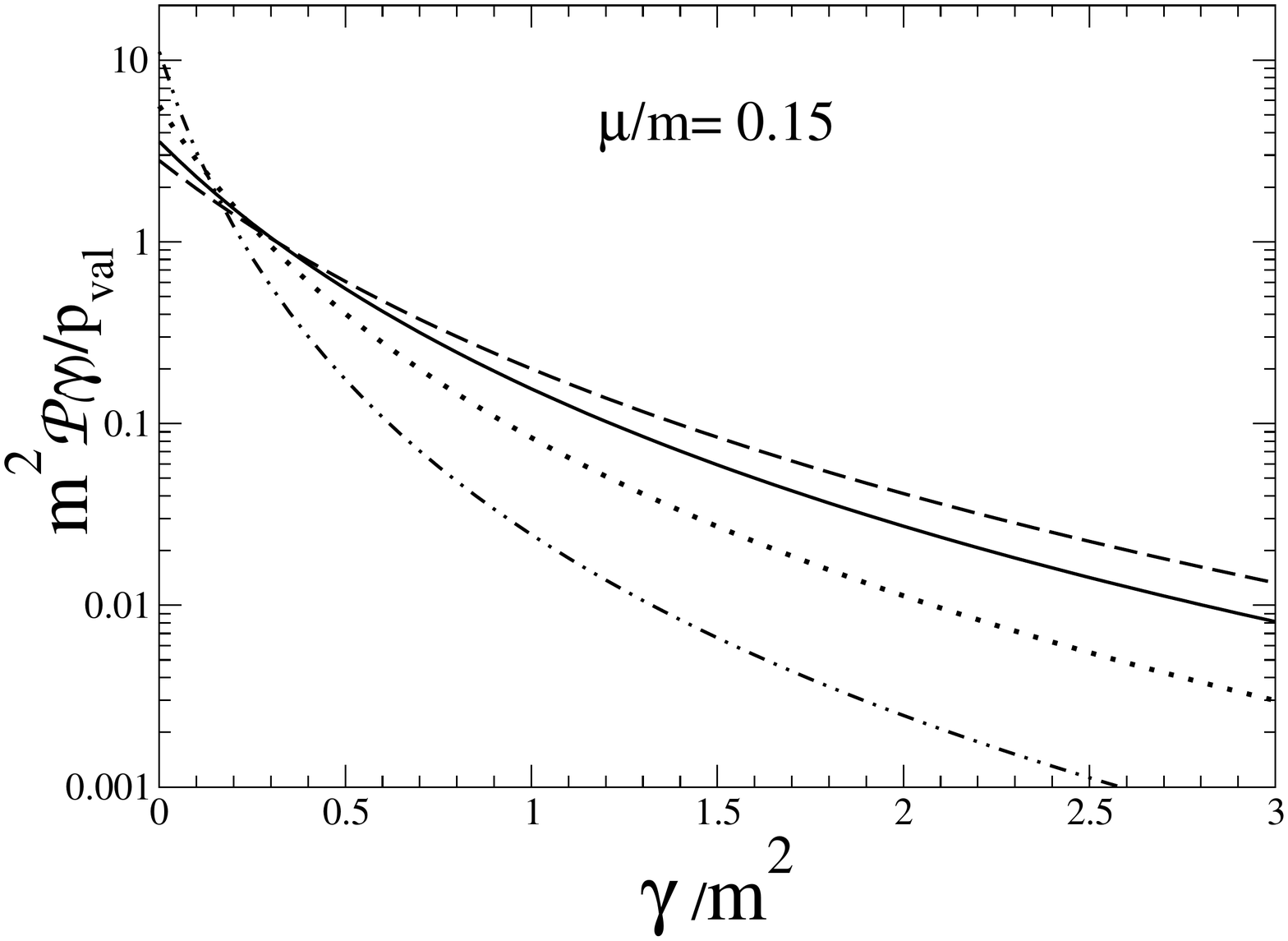}

\includegraphics[width=9.cm]{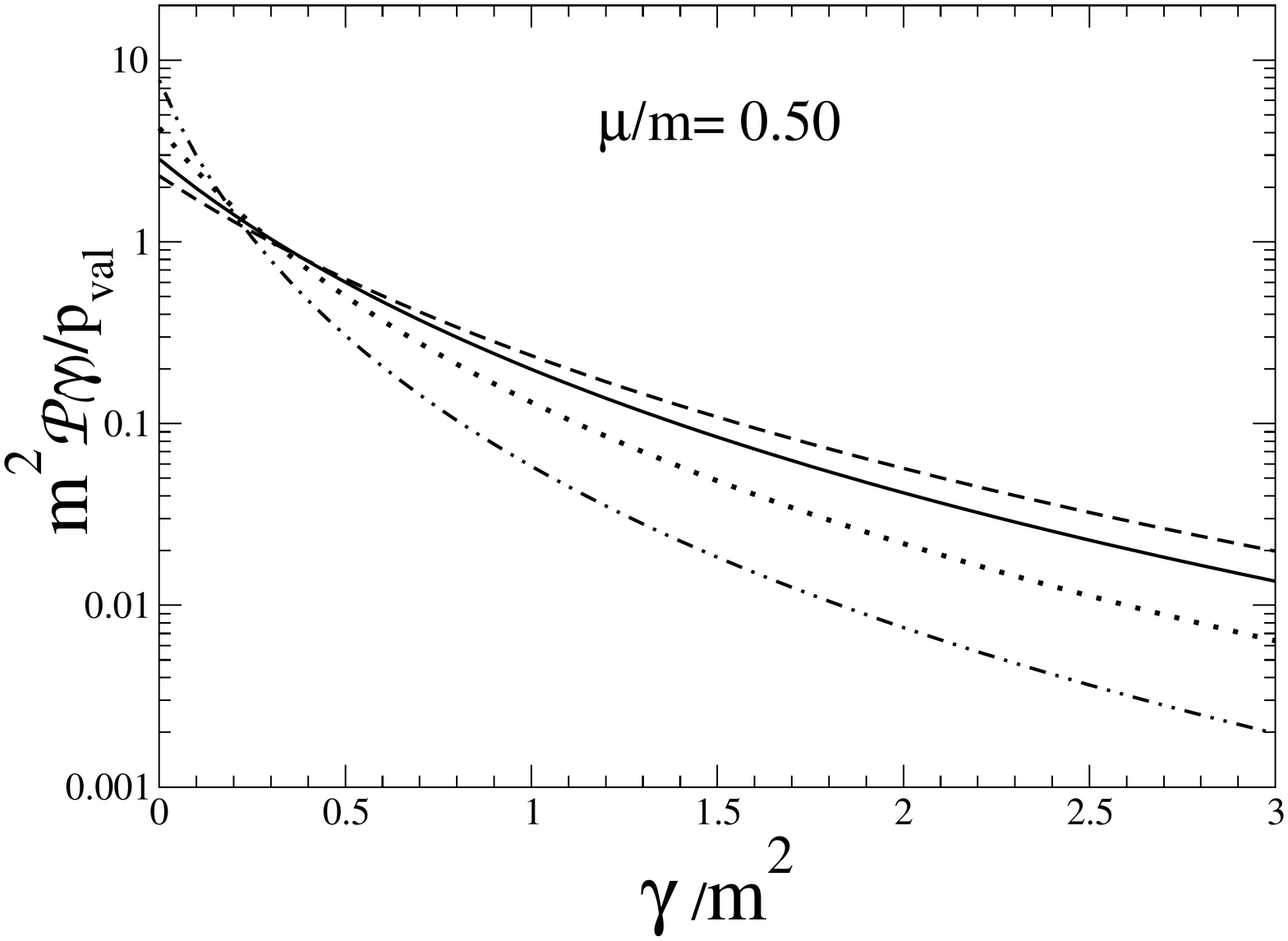}
\caption{The transverse LF-distribution ${\cal P}(\gamma)$ for the valence
component, Eq. (\ref{probgam}), vs the adimensional variable   $\gamma/m^2$, for 
$\mu/m=0.05,~0.15,~0.50$.   
Dash-double-dotted line: $B/m=0.20$. Dotted line: $B/m=0.50$. 
Solid line: $B/m=1.0$. Dashed line: $B/m=2.0$.
Recall that    $\gamma=k^2_\perp$  and $ \int_0^\infty d\gamma~{\cal P}(\gamma)=P_{val}$ (cf 
Table \ref{tab4}).
}
\label{pgam}
\end{figure}

In Figs. (\ref{fixi}) and (\ref{pgam}) the valence LF distributions, given in
Eqs. (\ref{phixi}) and (\ref{probgam}) are shown for $\mu/m=0.05,~0.15,~0.5$ and
$B/m=0.2,~0.5,~1.0, ~2.0$. The curves correspond to the eigen-vectors of Eq.
(\ref{newbound2}), since the ones obtained from Eq. (\ref{ptirl}) completely
overlap with  the previous ones, though the weight functions have differences at low values of $\gamma$.
  It should be pointed out that  
 the valence wave function, the main ingredient for
calculating the LF distributions, is obtained from the weight function by applying
the integral operator symbolically indicated by $B(M)$ in Eq. (\ref{symb}). This
eliminates
 the above mentioned instabilities, that  are possibly  produced  by
   the inversion of $B(M)$.

\begin{table}
\caption
{Values of $P_{val}$, Eq. (\ref{pval}), evaluated 
by using the weight function,
$g_b^{(Ld)}(\gamma,z;\kappa^2)$, corresponding to (\ref{newbound2}) (i.e. with the
application of the uniqueness theorem) are shown for  three values of $\mu/m$,
and  varying 
the  binding energy, $B/m$.Notice that for $B/m=0.001$ the values  $N_z=16$,
$N_g=48$ and $a=12$ have been
adopted  in Eqs. (\ref{bas1}) and  (\ref{bas3}), for obtaining a better
convergence.}

\parbox {5.1cm}{\begin{tabular}{c} 
$  \mu/m=0.05   $
\end{tabular} 
\begin{tabular}{|c|c|c|}
\hline
  B/m   &  $\alpha$ &   $P_{val}$ \\ \hline 
  0.001  &0.1685  & 0.94         \\ \hline 
  0.01   &0.3521  & 0.89   	\\ \hline
  0.10   & 1.191& 0.75  	  \\ \hline
  0.20   & 1.850& 0.72   	  \\ \hline
  0.50   & 3.358&0.68           \\ \hline
  1.00   & 5.056&0.66          \\ \hline
  2.00   & 6.336 &  0.65        \\ \hline
\end{tabular}}
\parbox {5.1cm}{\begin{tabular}{c} 
$  \mu/m=0.15   $
\end{tabular}
\begin{tabular}{|c|c|c|}
\hline
  B/m   &  $\alpha$ &   $P_{val}$ \\ \hline
  0.001 & 0.3667  & 0.97     \\ \hline  
  0.01  & 0.5716 & 0.94       \\ \hline 
  0.10  & 1.437 & 0.80   	  \\ \hline 
  0.20  & 2.099  & 0.75  	  \\ \hline 
  0.50  & 3.611 & 0.70      \\ \hline
  1.00  & 5.314 & 0.67       \\ \hline
  2.00  & 6.598 &  0.66      \\ \hline 
\end{tabular} }  
\parbox{5.0cm} {\begin{tabular}{c} 
$ \mu/m=0.50   $ \end{tabular}
\begin{tabular}{|c|c|c|}
\hline
  B/m   &  $\alpha$ &   $P_{val}$ \\ \hline 
  0.001 & 1.167 & 0.98    \\ \hline  
  0.01  & 1.440 & 0.96      \\ \hline 
  0.10  & 2.498 & 0.87   	  \\ \hline 
  0.20  & 3.251 & 0.83 	  \\ \hline 
  0.50  & 4.900 & 0.77      \\ \hline
  1.00  & 6.711 & 0.74       \\ \hline
  2.00  & 8.061 & 0.72     \\ \hline
\end{tabular}} 
\label{tab4}
\end{table}

\section{Conclusion}\label{concl}
We have quantitatively investigated the ladder Bethe-Salpeter Equation, in Minkowski space,
 within the perturbation-theory integral
representation of the multi-leg transition amplitudes, proposed by Nakanishi in the  60's \cite{nak63,nak71}.
The formal analysis leading to determine the Nakanishi weight function takes a 
great benefit from
the Light-Front framework, as shown in Ref. \cite{carbonell1} for the bound states and in Ref. \cite{FSV} for the
scattering states. In particular, if one exploit both i) the equation, obtained from BSE, 
for the valence component of the Fock expansion of
the interacting-system state and ii) the Nakanishi theorem  \cite{nak71} 
on the uniqueness of the non singular weight
function related to the  vertex function in PTIR, one can obtain Eq. (\ref{newbound2}).
This equation and the one in (\ref{ptirl}) allows the numerical evaluation of
 the weight function corresponding to a given value 
of the
binding energy of the interacting system and the exchanged-boson mass.
We have shown that the  eigenvalues  and the 
eigenvectors  obtained by solving
Eqs. (\ref{newbound2}) and (\ref{ptirl}) can be substantially taken  
as the same (the eigenvectors differ at the level of few-percent for $\gamma \to 0$).
In particular,  if one considers only the phenomenological observables, i.e. eigenvalues and LF distributions, 
the 
outcomes of
both equations can be even   taken as equal. 
This gives us great confidence in the validity of the uniqueness theorem also in a non 
perturbative regime.

An important feature of our analysis is represented by the
basis we have chosen for
expanding the weight function. Such a basis is able to include all the general properties of the 
weight function,
allowing a good control on the instabilities, that, in principle, could plague the numerical solutions of the 
above equations.

In perspective, the numerical analysis we have performed appears very encouraging, and it makes 
compelling
the next steps, 
represented by
the evaluation of observables related to the scattering states, like the scattering length, and 
the inclusion of
the crossed-box diagrams as already done in Ref. \cite{carbonell2}, but exploiting the Nakanishi uniqueness
theorem.
\section*{Acknowledgments}
We gratefully thank  Jaume
Carbonell and Vladimir Karmanov for very stimulating discussions.
TF  acknowledges the warm hospitality of INFN Sezione di Pisa and  thanks 
 the partial financial 
support from 
 the Conselho Nacional
de Desenvolvimento Cient\'{\i}fico e Tecnol\'ogico (CNPq),
the  Funda\c c\~ao de Amparo \`a Pesquisa do
Estado de S\~ao Paulo (FAPESP). 

\appendix
\section{A new form for the LF kernel in ladder approximation}
\label{nkern}
This Appendix contains the details for obtaining  the  expression 
of the kernel in Eq. (\ref{nkern1}), 
 ${ V}_b^{(Ld)}$,  that is suitable for applying the uniqueness
theorem for the Nakanishi weight function.

In a reference frame where ${\bf p}_\perp=0$ and $p^\pm=M$,
 the kernel in ladder approximation is (cf Ref. \cite{FSV})
\be
V^{(Ld)}_b(\gamma,z;\gamma',z')=
~-~g^2p^+
\int \frac{d^4k''}{(2\pi)^4}\frac{1}
{\left[{k''}^2+p\cdot k'' z'-\gamma'-\kappa^2+i\epsilon\right]^3}
~\times \nonu \int_{-\infty}^{\infty}{d k^- \over 2\pi}~
{1 \over \left[(\frac{p}{2}+k)^2-m^2+i\epsilon\right]}
~ {1\over \left[(\frac{p}{2}-k)^2-m^2+i\epsilon\right]}{1\over (k-k'')^2
-\mu^2 +i\epsilon}=
\nonu=
~{g^2 \over 2(4\pi)^2}~{1 \over \left[\gamma +(1-z^2)\kappa^2 +z^2 m^2
-i\epsilon\right]}
~
\int_0^1  d v ~v^2{\cal F}(v,\gamma,z;\gamma',\zeta')
\label{vlboundA}\ee
where
\be
{\cal F}(v,\gamma,z;\gamma',\zeta,\zeta')=
{(1+z)^2\over X^2(v,\zeta,\zeta')}~\
{\theta (\zeta'-z) \over \left[ \gamma +z^2 m^2+\kappa^2(1-z^2)+
\Gamma(v,z,\zeta',\gamma')
 -i\epsilon\right]^2}+ \nonu +
{(1-z)^2\over X^2(v,\zeta,-\zeta')} ~
{ \theta (z-\zeta') \over \left[\gamma +z^2 m^2+\kappa^2(1-z^2)+
\Gamma(v,-z,-\zeta',\gamma')
 -i\epsilon\right]^2}
\label{calf1}\ee
with
\be
X(v,\zeta,\zeta')=v(1-v)(1+\zeta')
\nonu
\Gamma(v,z,\zeta',\gamma')= { (1+z)\over (1+\zeta')}~\left\{{v\over (1-v)}\left[{\zeta'}^{2}
\frac{M^2}{4} + \kappa^2 (1+{\zeta}^2)+\gamma'\right]+{\mu^2\over v}
+\gamma'\right\}
\ee
The previous expression coincides with the one 
 in Ref. \cite{carbonell1}.  

For combining the denominators  in the last line of Eq. (\ref{vlboundA})
and in Eq. (\ref{calf1}) the standard Feynman trick can be used,
viz 
\be
{1\over B~A^2}= \lim_{\lambda \to 0^+}~{1 \over \lambda}
\left [{1 \over BA}~-{1\over B(A+ \lambda)}\right]=
\nonu=\lim_{\lambda \to 0^+}~{1 \over \lambda}
\left\{
\int_0^1 d\xi ~{1\over \left[ B-\xi (B-A) \right]^2}-
\int_0^1 d\xi ~{1\over \left[ B-\xi (B-A) +\xi\lambda\right]^2}\right\}
\label{newtri}\ee
with \be
A=\gamma +z^2 m^2+\kappa^2(1-z^2)+
\Gamma(v,\pm z,\pm \zeta',\gamma')
 -i\epsilon
 \nonu
 B=\gamma +z^2 m^2+\kappa^2(1-z^2)-i\epsilon
\ee
 obtaining 
 the following expression
\be
V^{(Ld)}_{b}(\gamma,z;\gamma',\zeta')= ~- ~
~{g^2 \over 2(4\pi)^2}~
 \left [{(1+z)\over (1+\zeta')}
~\theta (\zeta'-z)~{\cal H}'(\gamma,z;\gamma',\zeta',\mu^2)
+\right. \nonu \left . +{(1-z)\over (1-\zeta')}
~\theta
(z-\zeta')~{\cal H}'(\gamma,-z;\gamma',-\zeta',\mu^2)\right]
\ee
where
\be
{\cal H}'(\gamma,z;\gamma',\zeta'\mu^2)=
\lim_{\lambda \to 0^+}~{1 \over \lambda}
\left [ ~{\cal H}(\gamma,z;\gamma',\zeta',\mu^2,\lambda)
-{\cal H}(\gamma,z;\gamma',\zeta',\mu^2,0)\right] \label{a1}\ee
with 
\be{\cal H}(\gamma,z;\gamma',\zeta',\mu^2,\lambda)=
{(1+z)\over (1+\zeta')}
\int_0^1 {d v \over (1-v)^2}~\int_0^1 d\xi
\int^\infty_{-\infty}
 d\gamma'' \times \nonu
{\delta \left[\gamma'' - \xi \Gamma(v,z,\zeta',\gamma')
-\xi \lambda\right]\over \left[\gamma +\gamma''+z^2 m^2+
\kappa^2(1-z^2)
-i\epsilon   \right]^2} \ee
The positivity of  $\gamma'$ (cf Eq. (\ref{ptireq})) entails the positivity of 
$\Gamma(v,z,\zeta',\gamma')$ and, eventually,  of $\gamma''$. Given the linear dependence upon $\xi$ in
the delta function, one can productively perform first the integration on
 $\xi$, obtaining
\be
{\cal H}(\gamma,z;\gamma',\zeta',\mu^2,\lambda)=
 \int^\infty_{-\infty}
 d\gamma'' ~{
\theta(\gamma'')~h(\gamma'',z;\gamma',\zeta',\mu^2,\lambda)\over
\left[\gamma+\gamma'' +z^2 m^2+ \kappa^2(1-z^2)  -i\epsilon
\right]^2}  \ee
where
\be h(\gamma'',z;\gamma',\zeta',\mu^2,\lambda) = {(1+z)\over (1+\zeta')}
\int_0^1 {d v\over
(1-v)^2}\int_{0}^1 d\xi ~ \delta \left[\gamma'' - \xi
\Gamma(v,z,\zeta',\gamma') -\xi \lambda\right]=\nonu=
{(1+z)\over (1+\zeta')}\int_0^1 {d v\over (1-v)^2}{\theta\left(
\Gamma(v,z,\zeta',\gamma')-\gamma''  + \lambda
\right)\over \Gamma(v,z,\zeta',\gamma') + \lambda }
\label{k2} \ee
since $\xi$ must belong to the interval $[0,1]$. The derivative of 
$h(\gamma'',z;\gamma',\zeta',\mu^2,\lambda)$  implied by Eq. (\ref{a1}) is given
by
\be h'(\gamma'',z;\gamma',\zeta',\mu^2) ={(1+z)\over (1+\zeta')}
{1\over\gamma''}\int_0^1 {d v\over (1-v)^2}~\delta\left(
\Gamma(v,z,\zeta',\gamma')-\gamma''\right)+\nonu-{(1+z)\over (1+\zeta')}\int_0^1
{dv\over (1-v)^2}{\theta\left(
\Gamma(v,z,\zeta',\gamma')-\gamma'' \right)\over
[\Gamma(v,z,\zeta',\gamma')]^2 }=\nonu
 ={(1+z)\over (1+\zeta')}
{1\over\gamma''}\int_0^\infty {dy}~\delta\left(
\Gamma(y,z,\zeta',\gamma')-\gamma''\right) -{(1+z)\over (1+\zeta')}\int_0^\infty
{dy}~{\theta\left( \Gamma(y,z,\zeta',\gamma')-\gamma''
\right)\over [\Gamma(y,z,\zeta',\gamma')]^2 }.=
\nonu =
 {1\over\gamma''}\int_0^\infty {dy}~\delta\left[
{1 \over y}\left( y^2{\cal A}_b(\zeta',\gamma',\kappa^2 ) 
 +y {\cal B}_b(z,\zeta',\gamma', \gamma'',\mu^2 )   +\mu^2\right)\right] \nonu 
-~{ (1+\zeta')\over (1+z)}\int_0^\infty
{dy}~y^2~{\theta\left[y^2{\cal A}_b(\zeta',\gamma',\kappa^2 ) 
 +y {\cal B}_b(z,\zeta',\gamma', \gamma'',\mu^2 )+\mu^2\right]\over 
\left[ {y}^2{\cal A}_b(\zeta',\gamma',\kappa^2 )+ y(\mu^2 +\gamma')+\mu^2
\right ]^2 }. \label{k4} \ee
where the transformation $v\to y/(1+y)$ has been performed. In Eq. (\ref{k4})
 the function
 $\Gamma(y,z,\zeta',\gamma')$ is given by
 \be \Gamma(y,z,\zeta',\gamma')= { (1+z)\over
(1+\zeta')}~\left\{{y}\left[{\zeta'}^{2} \frac{M^2}{4} +
\kappa^2+\gamma'\right]+{ 1+y\over y}\mu^2 +\gamma'\right\}=
\nonu= { (1+z)\over
(1+\zeta')}~{1\over y}~\left\{y^2{\cal A}_b(\zeta',\gamma',\kappa^2 )+
y(\mu^2 +\gamma') +\mu^2\right\}\ee
and 
\be
{\cal A}_b(\zeta',\gamma',\kappa^2 )={\zeta'}^{2} \frac{M^2}{4} +
\kappa^2+\gamma'~\geq ~0
\nonu
{\cal B}_b(z,\zeta',\gamma', \gamma'',\mu^2 )=
\mu^2+\gamma' -\gamma''
{ (1+\zeta')\over (1+z)}
\ee
The two contributions to $h'(\gamma'',z;\gamma',\zeta',\mu^2)$ will be
discussed separately in what follows.
The first term is
\be
I_1={1\over\gamma''}\int_0^\infty {dy}~\delta\left[
{1 \over y}\left( y^2{\cal A}_b(\zeta',\gamma',\kappa^2 ) 
 +y {\cal B}_b(z,\zeta',\gamma', \gamma'',\mu^2 )   +\mu^2\right)\right]=
 \nonu=
 {1\over\gamma''} 
 \int_0^\infty {dy}~ \left [
 \theta(y_+)~y_+~
 {\delta(y-y_+)\over {\cal A}_b(\zeta',\gamma',\kappa^2)\left| y_+ - y_-\right|} +
 \theta(y_-)~y_-~{\delta(y-y_-)\over {\cal A}_b(\zeta',\gamma',\kappa^2) \left| y_+ - y_-\right|} \right]
 \nonu \times ~\theta(\Delta^2(z,\zeta',\gamma', \gamma'',\kappa^2,\mu^2 ) )
\label{i0}\ee
where $y_i$ are the two solutions of  
\be
y^2{\cal A}_b(\zeta',\gamma',\kappa^2 ) 
 +y {\cal B}_b(z,\zeta',\gamma', \gamma'',\mu^2 )   +\mu^2=0
\ee
namely
\be
y_\pm=
{1 \over 2{\cal A}_b(\zeta',\gamma',\kappa^2)} 
 \left[ -{\cal B}_b(z,\zeta',\gamma',\gamma'',\mu^2)
 \pm \Delta(z,\zeta',\gamma', \gamma'',\kappa^2,\mu^2 )\right]
 \label{i1}\ee
 with
 \be
\Delta^2(z,\zeta',\gamma', \gamma'',\kappa^2,\mu^2 )=
{\cal B}_b^2(z,\zeta',\gamma',\gamma'',\mu^2)
- 4 \mu^2~ {\cal A}_b(\zeta',\gamma',\kappa^2)~\geq ~0 
\ee
Notice that
\be
y_+~y_-=~{\mu^2\over {\cal A}_b(\zeta',\gamma',\kappa^2)}
\ee
This means that the two solutions have the same sign. Only for positive
solutions, $I_1$ is not vanishing.

From the requested positivity of the two solutions, one deduces that 
\be
0~ \geq ~{\cal B}_b(z,\zeta',\gamma',\gamma'',\mu^2)=\mu^2+\gamma' -\gamma''
{ (1+\zeta')\over
(1+z)} 
\ee 
Therefore the two constraints $\theta(y_+)$ and $\theta(y_-)$, once $y_{\pm}$ 
 exist,
are simultaneously fulfilled if
\be
\theta \left[\gamma'' { (1+\zeta')\over (1+z)} -\gamma' - \mu^2\right]
\ee
In conclusion, Eq. (\ref{i0}) becomes
\be I_1=
 {1\over\gamma''} 
 \theta \left[\gamma'' { (1+\zeta')\over (1+z)} -\gamma' - \mu^2\right]
  ~
 {y_+ + y_-\over {\cal A}_b(\zeta',\gamma',\kappa^2)\left| y_+ - y_-\right|} 
  ~\theta(\Delta^2(z,\zeta',\gamma', \gamma'',\kappa^2,\mu^2 ) )=
  \nonu=
 - ~{{\cal B}_b(z,\zeta',\gamma',\gamma'',\mu^2) \over
{\cal A}_b(\zeta',\gamma',\kappa^2) \Delta(z,\zeta',\gamma', \gamma'',
\kappa^2,\mu^2 ) }~{1\over\gamma''} 
 \theta \left[\gamma'' { (1+\zeta')\over (1+z)} -\gamma' - \mu^2\right]
\nonu \times~\theta(\Delta^2(z,\zeta',\gamma', \gamma'',\kappa^2,\mu^2 ) )
\label{i2}\ee
The second term in Eq. (\ref{k4}), i.e.
\be
I_2=-{ (1+\zeta')\over (1+z)}\int_0^\infty
{dy}~y^2~{\theta\left[y^2{\cal A}_b(\zeta',\gamma',\kappa^2 ) 
 +y {\cal B}_b(z,\zeta',\gamma', \gamma'',\mu^2 )+\mu^2\right]\over 
\left[ {y}^2{\cal A}_b(\zeta',\gamma',\kappa^2 )+ y(\mu^2 +\gamma')+\mu^2
\right ]^2 }
\ee
can be analyzed as follows.
One has to discuss two cases: i) if ${\cal B}_b(z,\zeta',\gamma', \gamma'',\mu^2
)\geq 0$  the argument of the theta function is positive for any $y$, ii) 
if ${\cal B}_b(z,\zeta',\gamma', \gamma'',\mu^2
) <0$ one has to check the sign of 
$\Delta^2(z,\zeta',\gamma', \gamma'',\kappa^2,\mu^2 )$. In conclusion, one can
single out the following three contributions  
\be
I^{(a)}_2=~-~{ (1+\zeta')\over (1+z)}
\theta({\cal B}_b(z,\zeta',\gamma', \gamma'',\mu^2 ))~
\int_0^{\infty}{dy}~{y^2 \over \left[ {y}^2{\cal A}_b(\zeta',\gamma',\kappa^2 )+ y(\mu^2 +\gamma')+\mu^2
\right ]^2}
\nonu
I^{(b)}_2=~-~{ (1+\zeta')\over (1+z)}\theta(-{\cal B}_b(z,\zeta',\gamma', \gamma'',\mu^2 ))~
\theta(\Delta^2(z,\zeta',\gamma', \gamma'',\kappa^2,\mu^2 ))
\nonu \times ~
\int_0^{\infty}{dy}~{y^2 \left[\theta(y-y_+) +\theta(y_- -y)\right]\over \left[ {y}^2{\cal A}_b(\zeta',\gamma',\kappa^2 )+ y(\mu^2 +\gamma')+\mu^2
\right ]^2}=
\nonu=~-~{ (1+\zeta')\over (1+z)}\theta(-{\cal B}_b(z,\zeta',\gamma', \gamma'',\mu^2 ))~
\theta(\Delta^2(z,\zeta',\gamma', \gamma'',\kappa^2,\mu^2 ))
\nonu \times ~
\int_0^{\infty}{dy}~{y^2 \left[1-\theta(y_+-y) +\theta(y_- -y)\right]\over \left[ {y}^2{\cal A}_b(\zeta',\gamma',\kappa^2 )+ y(\mu^2 +\gamma')+\mu^2
\right ]^2}
 \nonu
I^{(c)}_2=~-~{ (1+\zeta')\over (1+z)}
\theta(-{\cal B}_b(z,\zeta',\gamma', \gamma'',\mu^2 ))~
\theta(-\Delta^2(z,\zeta',\gamma', \gamma'',\kappa^2,\mu^2 ))
\nonu \times 
\int_0^{\infty}{dy}~{y^2 \over \left[ {y}^2{\cal A}_b(\zeta',\gamma',\kappa^2 )+ y(\mu^2 +\gamma')+\mu^2
\right ]^2}
\ee 

Therefore
\be
I_2=I^{(a)}_2+I^{(b)}_2+I^{(c)}_2
=-{ (1+\zeta')\over (1+z)}
\int_0^{\infty}{dy}~{y^2 \over \left[ {y}^2{\cal A}_b(\zeta',\gamma',\kappa^2 )+ y(\mu^2 +\gamma')+\mu^2
\right ]^2} \nonu 
+
{ (1+\zeta')\over (1+z)}
\theta( \gamma'' { (1+\zeta')\over (1+z)} -\gamma' - \mu^2)~
\theta(\Delta^2(z,\zeta',\gamma', \gamma'',\kappa^2,\mu^2 ))
\nonu \times ~
\int_{y_-}^{y_+}{dy}~{y^2 \over \left[ {y}^2{\cal A}_b(\zeta',\gamma',\kappa^2 )+ y(\mu^2 +\gamma')+\mu^2
\right ]^2}
\ee
Collecting the above results, Eq. (\ref{k4}) can be cast in the following form
\be
h'(\gamma'',z;\gamma',\zeta',\mu^2) 
=  \theta \left[\gamma'' { (1+\zeta')\over (1+z)} -\gamma' - \mu^2\right]
~\theta(\Delta^2(z,\zeta',\gamma', \gamma'',\kappa^2,\mu^2 ) ) 
\nonu \times\left\{- {{\cal B}_b(z,\zeta',\gamma',\gamma'',\mu^2) \over
\gamma''~{\cal A}_b(\zeta',\gamma',\kappa^2)~\Delta(z,\zeta',\gamma', \gamma'',\kappa^2,
\mu^2 ) }
 + \right. \nonu + \left.
{ (1+\zeta')\over (1+z)}~
\int_{y_-}^{y_+}{dy}~{y^2 \over \left[ {y}^2{\cal A}_b(\zeta',\gamma',\kappa^2 )+ y(\mu^2 +\gamma')+\mu^2
\right ]^2}\right\}
\nonu
-{ (1+\zeta')\over (1+z)}
\int_0^{\infty}{dy}~{y^2 \over \left[ {y}^2{\cal A}_b(\zeta',\gamma',\kappa^2 )+ y(\mu^2 +\gamma')+\mu^2
\right ]^2} 
\label{kfin}\ee

The two theta functions can be simplified taking profit of their interplay.
 
 The explicit form for $\Delta^2$ is
\be
\Delta^2(\pm z,\pm \zeta',\gamma', \gamma'',\kappa^2,\mu^2 )=
\left[\gamma''
{ (1 \pm \zeta')\over (1\pm z)}-\gamma'-\mu^2\right]^2 - 4 \mu^2~  \left({\zeta'}^{2} \frac{M^2}{4} +
\kappa^2+\gamma'\right)=
\nonu
= \left[\gamma''
{ (1 \pm \zeta')\over (1\pm z)}-\gamma'-\mu^2 - 2 \mu~  \sqrt{{\zeta'}^{2} \frac{M^2}{4} +
\kappa^2+\gamma'}\right]~\times \nonu 
\left[\gamma''
{ (1 \pm \zeta')\over (1\pm z)}-\gamma'-\mu^2 +2 \mu~  \sqrt{{\zeta'}^{2} \frac{M^2}{4} +
\kappa^2+\gamma'}\right]
\ee
In order to have 
$\Delta^2(\pm z,\pm \zeta',\gamma', \gamma'',\kappa^2,\mu^2 )\geq 0$, given the
presence of the first theta function in Eq. (\ref{kfin}), it is enough  that
 \be
 \gamma''
{ (1 \pm \zeta')\over (1\pm z)}-\gamma'-\mu^2 \geq 2 \mu~  \sqrt{{\zeta'}^{2} \frac{M^2}{4} +
\kappa^2+\gamma'}\geq 0
 \ee

Summarizing the above discussion, one can write the  kernel  as follows 
  \be
h'(\gamma'',z;\gamma',\zeta',\mu^2) =
\theta \left[\gamma'' { (1+\zeta')\over (1+z)} -\gamma' - \mu^2- 2\mu
\sqrt{{\zeta'}^{2} \frac{M^2}{4} + \kappa^2+\gamma'}~\right]
 \nonu \times
\left[- {{\cal B}_b(z,\zeta',\gamma',\gamma'',\mu^2) \over
{\cal A}_b(\zeta',\gamma',\kappa^2)~\Delta(z,\zeta',\gamma', \gamma'',\kappa^2,
\mu^2 ) }~{1\over\gamma''} \right. 
\nonu \left. + { (1+\zeta')\over (1+z)}
 \int_{y_-}^{y_+}{dy}~{y^2 \over \left[ {y}^2{\cal A}_b(\zeta',\gamma',\kappa^2 )+ y(\mu^2 +\gamma')+\mu^2
\right ]^2}\right]
\nonu
-{ (1+\zeta')\over (1+z)}
\int_0^{\infty}{dy}~{y^2 \over \left[ {y}^2{\cal A}_b(\zeta',\gamma',\kappa^2 )+ y(\mu^2 +\gamma')+\mu^2
\right ]^2} 
\label{kfin1}\ee

\section{The Normalization of the BS amplitude}
\label{snorm}
In this Appendix, the normalization of the BS amplitude, in ladder approximation, for a 
two-scalar system in
S-wave, is presented. The reader can find more details in Refs.   
\cite{lurie,zuber,nak63,nak69,nak88}.

In general, but disregarding self-energy contributions, the  BS amplitude is normalized through the following constraint
\be
 \int {d^4 k \over (2 \pi)^4}\int {d^4k'\over (2 \pi)^4}~\bar \Phi_b(k',p) ~\times \nonu
 \left.\left\{{\partial \over \partial
p^\mu} \left[ G^{-1}_0(12)(k,p)~(2\pi)^4~\delta^4(k-k') - i{\cal
K}(k,k',p)\right]\right\}\right|_{p^2=M^2} 
\!\Phi_b(k,p) = ~i~2 p_\mu
\label{nap1}\ee
with
\be
G_0(12)(k,p)=G_0\left({p\over 2}+k\right)~G_0\left({p\over 2}-k\right)=
{i \over ({p\over 2}+k)^2 -m^2 +i \epsilon}~{i \over ({p\over 2}-k)^2 -m^2 +i \epsilon}
\ee
In ladder approximation,  fortunately the kernel   $i{\cal K}(k,k',p)$ 
becomes independent of $p$, viz
\be
i{\cal K}^{(Ld)}(k,k',p)= {i (-i)^2\over (k-k')^2 -\mu^2 +i\epsilon}
\ee
 Therefore, the ladder BS amplitude is normalized through 
\be
~(-i)~\int {d^4 k \over (2 \pi)^4}~\bar \Phi^{(Ld)}_b(k,p) ~
\left[({p_\mu\over 2}+k_\mu) ~G^{-1}_0\left({p\over 2}-k\right)  +
 G^{-1}_0\left({p\over 2}+k\right)~({p_\mu\over 2}-k_\mu) \right]~ 
\Phi^{(Ld)}_b(k,p) = \nonu =~i~2 p_\mu
\label{nap2}\ee
Since $\Phi^{(Ld)}_b(k,p)$ is symmetric under the exchange $1 \to 2$, and recalling that $k^\mu$ changes sign under
such a transformation, one can rewrite Eq. (\ref{nap2}) as follows
\be
~-~\int {d^4 k \over (2 \pi)^4}~\bar \Phi^{(Ld)}_b(k,p) ~
(M^2+2 k\cdot p) ~\left[\left({p\over 2}-k\right)^2 -m^2  \right]~ 
\Phi^{(Ld)}_b(k,p) = \nonu=
\int {d^4 k \over (2 \pi)^4}~\bar \Phi^{(Ld)}_b(k,p) ~
\left[M^2 (\kappa^2-k^2) +2 (k\cdot p)^2\right] ~ 
\Phi^{(Ld)}_b(k,p)=~i~2 M^2
\ee
where $\kappa^2=m^2 -M^2/4$ and the odd contributions in $k^\mu$ have been eliminated, given the symmetry of the BS
amplitude.
By introducing the expression of $\Phi^{(Ld)}_b(k,p)$ in terms of the Nakanishi weight function, Eq. (\ref{naka1}),
with $n=1$ (as explained in the main text), one
gets 
\be
\int {d^4k\over (2 \pi)^4}\left[M^2 (\kappa^2-k^2) +2 (k\cdot p)^2\right]
~\nonu \times\int_{-1}^1 dz' \int_0^\infty d\gamma '{g^{(Ld)}_b(\gamma',z';\kappa^2)
\over [k^2 +p\cdot k z' -\gamma' -\kappa^2 +i \epsilon']^3} ~
\int_{-1}^1 dz \int_0^\infty d\gamma {g^{(Ld)}_b(\gamma,z;\kappa^2)
\over [k^2 +p\cdot k z -\gamma -\kappa^2 +i \epsilon]^3}= 
\nonu= \int_{-1}^1 dz' \int_0^\infty d\gamma '~g^{(Ld)}_b(\gamma',z';\kappa^2)
\int_{-1}^1 dz \int_0^\infty d\gamma ~g^{(Ld)}_b(\gamma,z;\kappa^2)~
{\cal F}(\gamma',z',\gamma,z)=
~i~2M^2
\label{nap3}\ee
 It is worth noting that  i) the S-wave  weight function 
is real, and ii)  the boundary condition $i\epsilon$ has to be chosen in  
$\bar \Phi_b(k,p)$
for ensuring the correct propagation in time (see, e.g.,
\cite{lurie,zuber}).

In order to evaluate ${\cal F}(\gamma',z',\gamma,z)$,  let us apply  the Feynman trick as follows
 (cf Refs. \cite{dae,carbonell4}
\be
{1\over [k^2 +p\cdot k z' -\gamma' -\kappa^2 +i \epsilon']^3}~
{1 \over [k^2 +p\cdot k z -\gamma -\kappa^2 +i \epsilon]^3}=\nonu =
\int ^1_0 dv ~{30 v^2(1-v)^2 \over \left[v
\left(k^2 +p\cdot k z' -\gamma' -\kappa^2 +i \epsilon'\right) +(1-v)\left( k^2 +p\cdot k z -\gamma -
\kappa^2 +i \epsilon \right)\right]^6}
=
\nonu = 
\int ^1_0 dv ~{30 v^2(1-v)^2 \over \left[ k^2 -\kappa^2 + p\cdot k~(vz' +(1-v)z) -\gamma' v
-\gamma(1-v) +i \eta\right]^6}=\nonu =
\int ^1_0 dv ~{30 v^2(1-v)^2 \over \left[
q^2 -\kappa^2 - {p^2 \over 4} \lambda^2  -\gamma' v
-\gamma(1-v) +i \eta\right]^6}
\ee
where $\eta=v\epsilon'+(1-v)\epsilon$,  and  
$ q=k + \lambda p/2$, with 
 $\lambda= \left[ vz' +(1-v)z\right]$. 
 By exploiting the above result, ${\cal F}(\gamma',z',\gamma,z)$ reduces to
 \be
 {\cal F}(\gamma',z',\gamma,z)=\int {d^4k \over (2 \pi)^4} {\left[M^2 (\kappa^2-k^2) +2 (k\cdot p)^2\right]\over [k^2 +p\cdot k z' -\gamma' -\kappa^2 +i \epsilon']^3}~
{1 \over [k^2 +p\cdot k z -\gamma -\kappa^2 +i \epsilon]^3}=
\nonu =
\int {d^4q \over (2 \pi)^4}~\int ^1_0 dv ~{\left [M^2 \left(\kappa^2 -q^2 +{M^2 \over
4}\lambda^2\right) +2 (p\cdot q)^2-M^2\lambda (p\cdot q)\right]30 v^2(1-v)^2 \over \left[
q^2 -\kappa^2 - {M^2 \over 4} \lambda^2  -\gamma' v
-\gamma(1-v) +i \eta\right]^6}=\nonu =
\int {d^4q \over (2 \pi)^4}~\int ^1_0 dv ~{\left [M^2 \left(\kappa^2 -q^2 +
{M^2 \over
4}\lambda^2\right) +2 (p\cdot q)^2\right]30 v^2(1-v)^2 \over \left[
q^2 -\kappa^2 - {M^2 \over 4} \lambda^2  -\gamma' v
-\gamma(1-v) +i \eta\right]^6}
 \ee
 where the term $(p\cdot q)$ yields a vanishing contribution after integrating
 over $d^4q$.
 
Then, by choosing  a reference frame where $p^\mu\equiv\{M,{\bf 0}\}$ for the
sake of simplicity, 
and performing a Wick rotation, given the positions of the poles, the
integration on $d^4 q$ can be evaluated in a Euclidean 4D space, i.e. $d^4 q \to
i d^4 q_E$. One obtains the
following result (cf   Refs. \cite{dae,carbonell4}) 
\be
{\cal F}(\gamma',z',\gamma,z)=\int {d^4q\over (2 \pi)^4} ~\int ^1_0 dv ~{\left [M^2 \left(\kappa^2 -q^2 +{M^2 \over
4}\lambda^2\right) +2 (p\cdot q)^2\right]30 v^2(1-v)^2 \over \left[
q^2 -\kappa^2 - {M^2 \over 4} \lambda^2  -\gamma' v
-\gamma(1-v) +i \eta\right]^6}=\nonu
=
~i~\int {d^4q_E\over (2 \pi)^4} ~\int ^1_0 dv ~{\left [M^2 \left(\kappa^2 +q^2_E +{M^2 \over
4}\lambda^2\right) -2 M^2 (q^0_E)^2\right]30 v^2(1-v)^2 \over \left[
q^2_E +\kappa^2 + {M^2 \over 4} \lambda^2  +\gamma' v
+\gamma(1-v) -i \eta\right]^6}=\nonu=~{iM^2\over (2 \pi)^4}~\int ^1_0 dv ~30 
v^2(1-v)^2
\int d\rho~\rho^3 ~\int_0^{2\pi} d\phi~
\int_{-1}^1 dcos
(\theta_1) \int_{0}^\pi sin^2(\theta_2)d
\theta_2 ~\times \nonu{\left [\kappa^2 +\rho^2 +{M^2 \over
4}\lambda^2 -2 \rho^2 cos^2(\theta_2)\right] \over \left[
\rho^2 +\kappa^2 + {M^2 \over 4} \lambda^2  +\gamma' v
+\gamma(1-v) -i \eta\right]^6}
\ee
with $$  d^4q_E \to   \rho^3 d\rho ~ d\phi~
 sin
(\theta_1)d\theta_1 ~ sin^2(\theta_2)d
\theta_2$$
Finally, by using
\be \int_0^{2\pi} d\phi~
\int_{-1}^1 dcos
(\theta_1) \int_{0}^\pi sin^2(\theta_2)d
\theta_2
\int_0^\infty d\rho ~{\rho^3 \over (\rho^2 +A)^6}= \nonu =
2\pi^2 ~{1 \over 2} \int_0^\infty dy~ {y \over (y +A)^6} = ~{ \pi^2 \over 20 ~A^4}
\ee
and
\be \int_0^{2\pi} d\phi~
\int_{-1}^1 dcos
(\theta_1) \int_{0}^\pi d
\theta_2~sin^2(\theta_2)~\left[1 -2 cos^2(\theta_2)\right]
\int_0^\infty d\rho ~{\rho^5~ 
\over (\rho^2 +A)^6}= \nonu=
 \pi^2~ {1\over 2}\int_0^\infty dy ~{y^2 \over (y +A)^6} = ~{ \pi^2 \over 60 ~A^3}
\ee
 one gets
 \be 
 {\cal F}(\gamma',z',\gamma,z)={i M^2\pi^2\over 2~(2 \pi)^4}~
 \int ^1_0 dv ~v^2(1-v)^2~\times \nonu
 {\left [3 \left(\kappa^2  +{M^2 \over
4}\lambda^2\right) +  \left( \kappa^2 + {M^2 \over 4} \lambda^2  +\gamma' v
+\gamma(1-v)    \right)\right]  \over \left[
\kappa^2 + {M^2 \over 4} \lambda^2  +\gamma' v
+\gamma(1-v) -i \eta\right]^4}
\ee
Recollecting the above results, the normalization condition, Eq. (\ref{nap3}), reads
\be
\int_{-1}^1 dz' \int_0^\infty d\gamma '~g^{(Ld)}_b(\gamma',z';\kappa^2)
\int_{-1}^1 dz \int_0^\infty d\gamma ~g^{(Ld)}_b(\gamma,z;\kappa^2)~
{\cal F}(\gamma',z',\gamma,z)=
\nonu={i M^2\over 2~(4 \pi)^2}~
\int_{-1}^1 dz' \int_0^\infty d\gamma '~g^{(Ld)}_b(\gamma',z';\kappa^2)
\int_{-1}^1 dz \int_0^\infty d\gamma ~g^{(Ld)}_b(\gamma,z;\kappa^2)~
 \int ^1_0 dv ~v^2(1-v)^2\times \nonu
 {\left [4 \left(\kappa^2  +{M^2 \over
4}\lambda^2\right) + \gamma' v
+\gamma(1-v)   \right]  \over \left[
\kappa^2 + {M^2 \over 4} \lambda^2  +\gamma' v
+\gamma(1-v) -i \eta\right]^4}=
~i~2M^2
\label{norfin}\ee
In conclusion, from Eq. (\ref{norfin}) one obtains the following 
 normalization of the Nakanishi weight function 
for the S-wave bound-state of a two-scalar
system 
\be
{1\over (8 \pi)^2}~
\int_{-1}^1 dz' \int_0^\infty d\gamma '~g^{(Ld)}_b(\gamma',z';\kappa^2)
\int_{-1}^1 dz \int_0^\infty d\gamma ~g^{(Ld)}_b(\gamma,z;\kappa^2)~
 \int ^1_0 dv ~v^2(1-v)^2\times \nonu
 {\left [4 \left(\kappa^2  +{M^2 \over
4}\lambda^2\right) +  \gamma' v
+\gamma(1-v)   \right]  \over \left[
\kappa^2 + {M^2 \over 4} \lambda^2  +\gamma' v
+\gamma(1-v) -i \eta\right]^4}=
~1
\label{normg}\ee



\begin{thebibliography}{99}
\bibitem{SB_PR84_51}
E.E.~Salpeter, H.A.~Bethe, Phys. Rev. {\bf 84}, 1232 (1951).
\bibitem{KW1} K.~Kusaka, A.G.~Williams, Phys. Rev. {\bf D 51}, 7026 (1995).

\bibitem{KW} K.~Kusaka, K.~Simpson, A.G.~Williams, Phys. Rev. {\bf D 56}, 5071
(1997).

\bibitem{dse} V. Sauli and J. Adam, Phys. Rev  {\bf D 67}, 085007 (2003).
\bibitem{sauli} V. Sauli,
Jou.  Phys. {\bf  G 35}, 035005 (2008); Few-Body Sys. {\bf 39}, 45 (2006).
\bibitem{carbonell1} J. Carbonell , V.A. Karmanov, Eur. Phys. J. {\bf A27}, 1 (2006).
 \bibitem{carbonell2} J. Carbonell , V.A. Karmanov, Eur. Phys. J. {\bf A27}, 11 (2006).
\bibitem{carbonell3} J. Carbonell, V.A. Karmanov Eur. Phys. J. {\bf A46}, 387 (2010).
\bibitem{nak63}
N.~Nakanishi, Phys. Rev. {\bf 130}, 1230 (1963).
\bibitem{nak69} N.~Nakanishi,
Prog. Theor. Phys. Suppl. {\bf 43}, 1 (1969).
\bibitem{nak88} N.~Nakanishi,
 Prog. Theor. Phys. Suppl. 
{\bf 95}, 1 (1988).
\bibitem{nak71} N.~Nakanishi, {\it Graph Theory
and Feynman Integrals} (Gordon and Breach, New York, 1971).
\bibitem{WICK_54} G.C.~Wick,     Phys. Rev. {\bf 96}, 1124 (1954).
\bibitem{CUTK_54} R.E.~Cutkosky, Phys. Rev. {\bf 96}, 1135 (1954).
\bibitem{cdkm} J.~Carbonell, B.~Desplanques, V.A.~Karmanov and
J.F.~Mathiot, Phys. Reports, {\bf 300}, 215 (1998).
\bibitem{Yabuki} H. Yabuki, Prog. Theor. Phys. {\bf 58}, 1935 (1977).
\bibitem{Burov} S. G. Bondarenko, V. V. Burov, A. V. Molochkov, G. I. Smirnov
and H. Toki, Prog.  Part. Nucl. Phys. {\bf 48}, 449 (2002).
\bibitem{FSV} T. Frederico, G. Salm\`e and M. Viviani, Phys. Rev. {\bf D 85},
036009 (2012).
\bibitem{sales00}J. H. O. Sales, T. Frederico, B. V. Carlson, P.
U. Sauer, Phys. Rev. {\bf C 61}, 044003 (2000).

\bibitem{sales01}J. H. O. Sales, T. Frederico, B. V. Carlson, P.
U. Sauer, Phys. Rev. {\bf C 63}, 064003 (2001).

\bibitem{hierareq}T. Frederico, J. H. O. Sales, B.V. Carlson and P.U. Sauer
Nucl. Phys. {\bf A 737}, 260c (2004).

\bibitem{adnei07} J. A. O. Marinho, T. Frederico, P. U. Sauer,
Phys. Rev. {\bf D 76}, 096001 (2007).
\bibitem{adnei08} J.A.O. Marinho, T. Frederico, E. Pace, G. Salme, P. U. Sauer,  Phys. Rev.
{\bf D 77}, 116010 (2008).
\bibitem{Sawicki} M. Sawicki, Phys. Rev. {\bf D 44}, 433 (1991).
\bibitem{carbonell5} J. Carbonell and V. A. Karmanov, Few-body Syst. {\bf 49},
205 (2011).


\bibitem{Brodrev} S.J. Brodsky, H.C. Pauli and S.S. Pinsky,
  Phys.Rep. {\bf 301}, 299 (1998).
  \bibitem{nak64} N. Nakanishi, Phys. Rev. {\bf 133}, B214
(1964).
\bibitem{nak642} N. Nakanishi,  Phys. Rev. {\bf 135}, B1224 (1964).
\bibitem{dae} Dae Sung Hwang and V. A. Karmanov, Nucl. Phys. {\bf 696}, 413 (2004).

\bibitem{lurie} D.Luri\'e, A. J.  Macfarlane and  Y. Takahashi, Phys. Rev. {\bf 140}, B1091 
(1965). 
\bibitem{zuber} C. Itzykson, J.B. Zuber "Quantum Field Theory",
 Dover Publications (2006).

\bibitem{gbaym} G. Baym,  Phys. Rev. {\bf 117}, 886 (1959).
\bibitem{gross} F. Gross,  \c{C}.  \c{S}avkli and J. Tjon, Phys. Rev.  {\bf D 64}, 
076008 (2001).
\bibitem{LM}  E. zur Linden and H. Mitter, Nuovo Cimento {\bf B 61}, 389 (1969).
\bibitem{carbonell7}  M. Mangin-Brinet and J. Carbonell, Phys. lett. {\bf B 474}, 237
(2000).
\bibitem{maris}  V.A. Karmanov and P. Maris, Few-body Syst. {\bf 46},95 (2009).

\bibitem{carbonell4}J. Carbonell, V.A. Karmanov, M. Mangin-Brinet,
Eur. Phys. J. {\bf A39}, 53 (2009).

\end{thebibliography}
\end{document}